\documentclass[11pt]{article}
\usepackage[margin=1in]{geometry}
\usepackage{titlepic}
\usepackage{graphicx}
\usepackage{float}
\usepackage{times}
\usepackage{url}
\usepackage[shortlabels]{enumitem}
\usepackage[table]{xcolor}
\usepackage{csquotes}
\usepackage{multicol}
\usepackage{titlesec}
\usepackage[numbers,square]{natbib}
\usepackage[bitstream-charter]{mathdesign}
\usepackage[T1]{fontenc}
\usepackage{setspace}
\usepackage{multirow}
\usepackage{array}
\usepackage[font={normal,it},labelfont=bf]{caption}
\usepackage{subcaption}
\usepackage{booktabs}
\usepackage{longtable}
\usepackage{changepage}
\usepackage{textcomp}
\usepackage{ragged2e}
\usepackage{amsmath}
\usepackage{xurl} 
\usepackage{hyperref}
\usepackage[nameinlink]{cleveref}
\usepackage[titletoc]{appendix}
\usepackage{afterpage}
\usepackage{microtype}
\usepackage{newfloat}
\usepackage{etoolbox}
\usepackage{tocloft}

\makeatletter
\renewcommand\@makefntext[1]{%
    \noindent
    \hb@xt@1.8em{\hss\@makefnmark\kern0.15em}#1}
\makeatother

\DeclareFloatingEnvironment[
    fileext=pdf,
    listname={Scenarios},
    name=Scenario,
    placement=tbhp,
]{scenario}
\Crefname{scenario}{Scenario}{Scenarios}
\crefname{scenario}{Scenario}{Scenarios}

\usepackage{tocloft}
\renewcommand\cftsecfont{\vskip-5pt}
\renewcommand\thesection{\arabic{section}}

\usepackage{titlesec}
\titleformat{\section}
  {\LARGE}{\thesection}{1em}{}
\usepackage{titlesec}
\titleformat{\subsection}
  {\Large}{\thesubsection}{1em}{}

\renewcommand\thesection{\arabic{section}}
\renewcommand{\thesubsection}{\thesection.\arabic{subsection}}
\makeatletter
\def\blfootnote{\xdef\@thefnmark{}\@footnotetext}
\makeatother

\crefformat{section}{\textbf{#2Section~#1#3}}
\Crefformat{section}{\textbf{#2Section~#1#3}}

\crefmultiformat{section}{\textbf{#2Sections~#1#3}}{ and \textbf{#2#1#3}}{, \textbf{#2#1#3}}{, and \textbf{#2#1#3}}
\Crefmultiformat{section}{\textbf{#2Sections~#1#3}}{ and \textbf{#2#1#3}}{, \textbf{#2#1#3}}{, and \textbf{#2#1#3}}

\crefmultiformat{appendix}{\textbf{#2Appendices~#1#3}}{ and \textbf{#2#1#3}}{, \textbf{#2#1#3}}{, and \textbf{#2#1#3}}
\Crefmultiformat{appendix}{\textbf{#2Appendices~#1#3}}{ and \textbf{#2#1#3}}{, \textbf{#2#1#3}}{, and \textbf{#2#1#3}}

\titlespacing*{\subsubsection}{0pt}{0.25\baselineskip}{0.25\baselineskip}
\titlespacing*{\subsection}{0pt}{0.4\baselineskip}{0.4\baselineskip}

\makeatletter
\newcommand{\appsection}[1]{
  \refstepcounter{section} 
  \addcontentsline{toc}{appsec}{\protect\numberline{\thesection}#1} 
  \section*{\thesection \quad #1} 
}

\let\l@appsec\l@section
\patchcmd{\l@appsec}{\cftsecfont}{\cftappsecfont}{}{}
\patchcmd{\l@appsec}{#2}{\cftappsecpagefont #2}{}{}

\apptocmd{\l@appsec}{\setlength{\cftbeforesecskip}{.1ex}}{}{}

\patchcmd{\l@appsec}{\numberline{}{#1}}{\numberline{#1}}{}{}

\makeatother
\newcommand{\cftappsecfont}{\cftsecfont}
\newcommand{\cftappsecpagefont}{\cftsecpagefont}

\crefname{appsec}{Appendix}{Appendices}
\crefname{appsection}{Appendix}{Appendices}
\creflabelformat{appsec}{\textbf{#2Appendix~#1#3}}
\renewcommand\thesubsection{\thesection.\arabic{subsection}}

\begin{document}

\begin{titlepage}

\newgeometry{margin=1.5cm}

\title{\huge Personhood credentials: \\ Artificial intelligence and the value of  \\ privacy-preserving tools to distinguish who is real online \blfootnote{$^{\dag}$ Indicates the corresponding authors: Steven Adler (steven\_adler@alumni.brown.edu), Zoë Hitzig (zhitzig@openai.com), and Shrey Jain (shreyjain@microsoft.com).} \blfootnote{$^{\ast}$ Denotes primary authors, who contributed most significantly to the direction and content of the paper. Besides corresponding authors, all other authors are listed in alphabetical order.}}

\author{\small Steven Adler,$^{\ast \dag 1}$ Zoë Hitzig,$^{\ast \dag 1,2}$ Shrey Jain,$^{\ast \dag 3}$ Catherine Brewer,$^{\ast 4}$ Wayne Chang,$^{\ast 5}$ Ren{é}e DiResta,$^{\ast 25}$ Eddy Lazzarin,$^{\ast 6}$ \\ \small Sean McGregor,$^{\ast 7}$ Wendy Seltzer,$^{\ast 8}$ Divya Siddarth,$^{\ast 9}$ Nouran Soliman,$^{\ast 10}$ Tobin South,$^{\ast 10}$ Connor Spelliscy,$^{\ast 11}$ \\ \small Manu Sporny,$^{\ast 12}$ Varya Srivastava,$^{\ast 4}$ John Bailey,$^{13}$ Brian Christian,$^{4}$ Andrew Critch,$^{14}$ Ronnie Falcon,$^{15}$ Heather Flanagan,$^{25}$ \\ \small Kim Hamilton Duffy,$^{16}$ Eric Ho,$^{17}$ Claire R. Leibowicz,$^{18}$ Srikanth Nadhamuni,$^{19}$ Alan Z. Rozenshtein,$^{20}$ \\ \small David Schnurr,$^{1}$ Evan Shapiro,$^{21}$ Lacey Strahm,$^{15}$ Andrew Trask,$^{4,15}$ Zoe Weinberg,$^{22}$ Cedric Whitney,$^{23}$ Tom Zick$^{24}$ 
\vspace{0.3in}\\
\footnotesize{$^1$OpenAI, $^2$Harvard Society of Fellows, $^3$Microsoft, $^4$University of Oxford, $^5$SpruceID, $^6$a16z crypto,}\\
\footnotesize{$^7$UL Research Institutes, $^8$Tucows, $^{9}$Collective Intelligence Project, $^{10}$Massachusetts Institute of Technology,}\\
\footnotesize{$^{11}$Decentralization Research Center, $^{12}$Digital Bazaar, $^{13}$American Enterprise Institute, }\\ 
\footnotesize{$^{14}$Center for Human-Compatible AI, University of California, Berkeley, $^{15}$OpenMined,}\\
\footnotesize{$^{16}$Decentralized Identity Foundation, $^{17}$Goodfire, $^{18}$Partnership on AI, $^{19}$eGovernments Foundation,}\\
\footnotesize{$^{20}$University of Minnesota Law School, $^{21}$Mina Foundation, $^{22}$ex/ante, $^{23}$School of Information, University of California, Berkeley,}\\
\footnotesize{ $^{24}$Berkman Klein Center for Internet \& Society, Harvard University, $^{25}$Independent Researcher}}

\renewcommand*{\thefootnote}{\fnsymbol{footnote}}
\setcounter{footnote}{0}

\date{August 2024}
\maketitle

\thispagestyle{empty}

\begin{abstract}\noindent \small Anonymity is an important principle online. However, malicious actors have long used misleading identities to conduct fraud, spread disinformation, and carry out other deceptive schemes. With the advent of increasingly capable AI, bad actors can amplify the potential scale and effectiveness of their operations, intensifying the challenge of balancing anonymity and trustworthiness online. In this paper, we analyze the value of a new tool to address this challenge: ``personhood credentials'' (PHCs), digital credentials that empower users to demonstrate that they are real people—not AIs—to online services, without disclosing any personal information. Such credentials can be issued by a range of trusted institutions---governments or otherwise. A PHC system, according to our definition, could be local or global, and does not need to be biometrics-based. Two trends in AI contribute to the urgency of the challenge: AI’s increasing indistinguishability from people online (i.e., lifelike content and avatars, agentic activity), and AI’s increasing scalability (i.e., cost-effectiveness, accessibility). Drawing on a long history of research into anonymous credentials and ``proof-of-personhood'' systems, personhood credentials give people a way to signal their trustworthiness on online platforms, and offer service providers new tools for reducing misuse by bad actors. In contrast, existing countermeasures to automated deception---such as CAPTCHAs---are inadequate against sophisticated AI, while stringent identity verification solutions are insufficiently private for many use-cases. After surveying the benefits of personhood credentials, we also examine deployment risks and design challenges. We conclude with actionable next steps for policymakers, technologists, and standards bodies to consider in consultation with the public.
\end{abstract}

\end{titlepage}

\restoregeometry

\renewcommand*{\thefootnote}{\arabic{footnote}}
\setcounter{footnote}{0}
\setcounter{page}{2}

\section*{Executive Summary}
\label{sec:executive}
\addcontentsline{toc}{section}{Executive Summary}

\small

\textbf{Malicious actors have long used misleading identities to deceive others online.} They carry out fraud, cyberattacks, and disinformation campaigns from multiple online aliases, email addresses, and phone numbers. Historically, such deception has sometimes seemed an unfortunate but necessary cost of preserving the Internet’s commitments to privacy and unrestricted access. But highly capable AI systems may change the landscape: There is a substantial risk that, without further mitigations, deceptive AI-powered activity could overwhelm the Internet. To uphold user privacy while protecting against AI-powered deception, new countermeasures are needed.

\textbf{With access to increasingly capable AI, malicious actors can potentially orchestrate more effective deceptive schemes.} Two trends contribute to these schemes' potential impact:
\begin{enumerate}[noitemsep, topsep=0pt]
    \item \textbf{Indistinguishability.} Distinguishing AI-powered users on the Internet is becoming increasingly difficult, as AI advances in its ability to:
    \begin{itemize}[noitemsep, topsep=0pt]
        \item \underline{Generate human-like content} that expresses human-like experiences or points of view (e.g., ``Here is what I thought of that speech'').
        \item \underline{Create human-like avatars} through photos, videos, and audio (e.g., simulating a real-looking person on a video chat).
        \item \underline{Take human-like actions} across the Internet (e.g., browsing websites like an ordinary user, making sophisticated plans to achieve goals they are given, solving CAPTCHAs when challenged).
    \end{itemize}
    \vspace{2mm}
    \item \textbf{Scalability.} AI-powered deception by malicious actors is increasingly scalable because of:
    \begin{itemize}[noitemsep,topsep=0pt]
        \item \underline{Decreasing costs} at all capability levels.
        \item \underline{Increasing accessibility}, for example, via open-weights deployments through which scaled misuse is less preventable.
    \end{itemize}
\end{enumerate}
Taken together, these two trends suggest that AI may help to make deceptive activity more convincing (through increased indistinguishability) and easier to carry out (through increased scalability).

\textbf{We identify one promising solution to pervasive deception on the Internet, building off decades of research in cryptography and experimentation in online communities: personhood credentials} (hereafter referred to as PHCs). Such a credential empowers its holder to demonstrate to providers of digital services that they are a person without revealing anything further. Building on related concepts like proof-of-personhood and anonymous credentials, these credentials can be stored digitally on holders’ devices and  verified through zero-knowledge proofs. Importantly, such proofs do not reveal the individual's specific credential (nor any aspects of their identity).

\textbf{To counter scalable deception while maintaining user privacy, PHC systems must meet two foundational requirements:}
\begin{enumerate}[noitemsep,topsep=0pt]
    \item \underline{Credential limits}: The issuer of a PHC gives at most one credential to an eligible person.
    \item \underline{Unlinkable pseudonymity}: PHCs let a user interact with services anonymously through a service-specific pseudonym; the user’s digital activity is untraceable by the issuer and unlinkable across service providers, even if service providers and issuers collude.
\end{enumerate}
These two properties equip service providers with the option to offer services on a per-person basis, and to prevent the return of users who violate the service's rules. An anonymous forum, for instance, could offer a single verified account to each credential holder. Unlinkable pseudonymity helps them achieve this because it prevents one person from using the same PHC to sign up twice, even without ever identifying the user. The issuer's credential limit gives them high confidence that the same user cannot easily circumvent the limit by using many PHCs to make many different accounts.

\textbf{There are many effective ways to design a PHC system, and various organizations---governmental or otherwise---can serve as issuers.} In one possible implementation, states could offer a PHC to any holder of their state’s tax identification number; a PHC system, according to our definition, could be local or global, and does not need to be based in biometrics. Having multiple trusted PHC issuers within a single ecosystem promotes choice---people can select into systems  built on their preferred root of trust (government IDs, social graphs, biometrics) and that offer affordances that best align with their preferences. This approach reduces the risks associated with a single centralized issuer while still preserving the ecosystem’s integrity by limiting the total number of credentials. Note that this paper does not advocate for or against any specific PHC system design; instead, it aims to establish the value of PHCs in general while highlighting challenges that must be taken into account in any design.

\textbf{PHCs are not forgeable by AI systems, and it is difficult for malicious actors to obtain many of them.} By combining verification techniques that have an offline component (e.g., appearing in-person, validating a physical document) and secure cryptography, these credentials are issued only to people and cannot be convincingly faked thereafter. They therefore help to counter the problem of indistinguishability by creating a credential only people can acquire, and help to counter the problem of scalability by enabling per-credential rate limits on activities.

\textbf{PHCs give digital services a tool to reduce the efficacy and prevalence of deception,} especially in the form of:
\begin{enumerate}[noitemsep,topsep=0pt]
    \item \underline{Sockpuppets}: deceptive actors purporting to be ``people'' that do not actually exist.
    \item \underline{Bot attacks}: networks of bots controlled by malicious actors to carry out automated abuse (e.g., breaking site rules and evading suspension by creating new accounts). 
    \item \underline{Misleading agents}: AI agents misrepresenting whose goals they serve.
\end{enumerate}
PHCs offer people a tool to credibly signal that they are a real person operating an authentic account, without conveying their identity. PHCs also help service providers spot deceptive accounts, which may lack such a signal. 

\textbf{PHCs improve on and complement existing approaches to countering AI-powered deception online.} For example, the following approaches are often not robust to highly capable AI, not inclusive, and/or not privacy-preserving:
\begin{enumerate}[noitemsep,topsep=0pt]
    \item \underline{Behavioral filters}, e.g., CAPTCHAs, JavaScript browser challenges, anomaly detection.
    \item \underline{Economic barriers}, e.g., paid subscriptions, credit card verification.
    \item \underline{AI content detection}, e.g., watermarking, fingerprinting, metadata provenance. 
    \item \underline{Appearance- and document-based verification}, e.g., selfie checks with ID, live video calls.
    \item \underline{Digital and hardware identifiers}, e.g., phone numbers, email addresses, hardware security keys.
\end{enumerate}

To achieve their benefits, \textbf{PHC systems must be designed and implemented with care.} We discuss four areas in which PHCs' impacts must be carefully managed:
\begin{enumerate}[noitemsep,topsep=0pt]
    \item \underline{Equitable access} to digital services that use PHCs.
    \item \underline{Free expression} supported by confidence in the privacy of PHCs.
    \item \underline{Checks on power} of service providers and PHC issuers.
    \item \underline{Robustness to attack and error} by different actors in the PHC ecosystem.
\end{enumerate}

\textbf{In close collaboration with the public, we encourage governments, technologists, and standards bodies to invest in the development, piloting, and adoption of personhood credentials} as a key tool in addressing scalable deception online:
\begin{enumerate}[noitemsep,topsep=0pt]
    \item \underline{Invest in development} and piloting of personhood credentialing systems. \\
    \hspace{\parindent} e.g., explore building PHCs incrementally atop existing credentials such as digital driver’s licenses.
    \item \underline{Encourage adoption} of personhood credentials. \\
    \hspace{\parindent} e.g., determine services for which PHCs ought to be substitutable for ID verification.
\end{enumerate}

\textbf{It is also important that these groups accelerate their preparations for AI’s impact more generally by adapting existing digital systems}:
\begin{enumerate}[noitemsep,topsep=0pt]
    \item \underline{Reexamine standards} for remote identity verification and authentication. \\
    \hspace{\parindent} e.g., reconsider confidence in selfie-based identity verification, absent supplemental factors to reduce AI-enabled spoofing.
    \item \underline{Study the impact and prevalence} of deceptive accounts across major communications platforms. \\
    \hspace{\parindent} e.g., develop standardized methods for measuring the prevalence of fake accounts on social media.
    \item \underline{Establish norms and standards} to govern agentic AI users of the Internet. \\
    \hspace{\parindent} e.g., explore new forms of trust infrastructure for AI agents, akin to HTTPS for websites.
\end{enumerate}
Readers primarily interested in ideas for next steps may refer directly to \Cref{sec:next_steps} for further detail.

\textbf{We are concerned that the Internet is inadequately prepared for the challenges highly capable AI may pose.} Without proactive initiatives involving the public, governments, technologists, and standards bodies, there is a significant risk that digital institutions will be unprepared for a time when AI-powered agents, including those leveraged by malicious actors, overwhelm other activity online. Lacking better alternatives, institutions might resort to privacy-violating methods for rooting out scaled deception, like creating digital identification systems that (intentionally or unintentionally) link a person’s legal identity with a complete record of their digital activity. By contrast, personhood credentials have the potential to reduce deceptive activity while preserving privacy---giving people and services the tools to signal and sustain trustworthiness online.

\begin{figure}[ht]
    \centerline{\includegraphics[width=.9\linewidth]{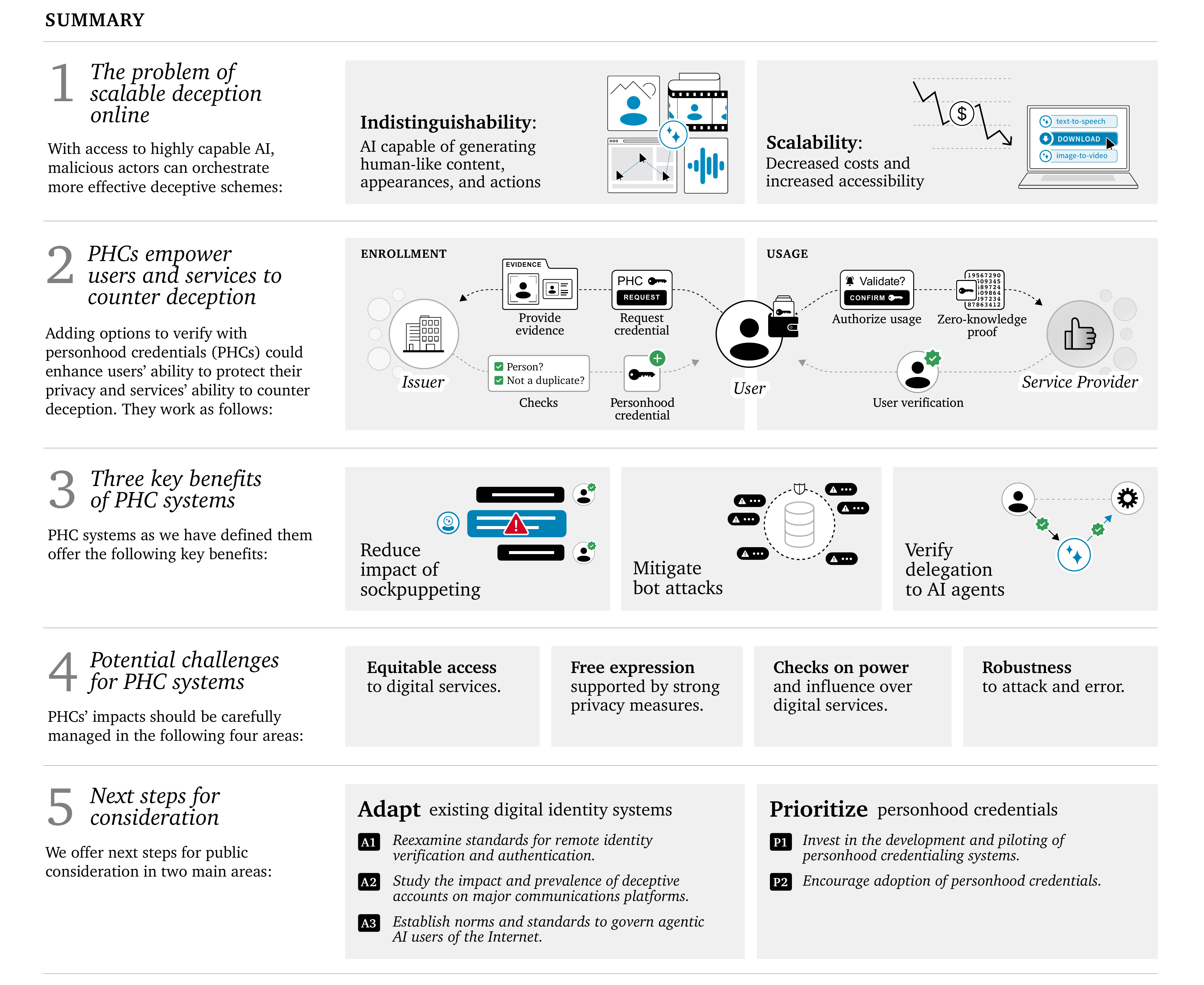}}
    \label{fig:exec_summary}
\end{figure}

\tableofcontents
\clearpage

\section{Introduction}\label{sec:introduction}

\subsection{The Internet has long struggled with deceptive activity}
\label{ssec:struggle_deception}

\normalsize 

Malicious actors have long used misleading identities to carry out abuse online. For example, attackers manipulate perceptions of public opinion by spreading disinformation through deceptive ``sockpuppet'' accounts—appearing to represent distinct people and thus lending their claims more credibility.\footnote{For an overview of the prevalence of these attacks, see \cite{bradshaw_industrialized_2021}. For an influential early attempt to detect astroturfing on Twitter, see \cite{ratkiewicz_truthy_2011}.} Other deceptive attacks employ automated botnets, executing distributed attacks by posing as a large number of distinct users.\footnote{A ``botnet'' originally referred to a network of compromised hardware devices, which could be controlled together. With the advent of cloud computing, botnets can also be run from rented computers in the cloud rather than relying on infected host computers. Botnets are used to carry out DDoS and credential stuffing attacks, to spread spyware, and to propagate scams \cite{cooke_zombie_2005, silva_botnets_2013}.} In a wide range of digital systems, the ability to create deceptive and seemingly independent profiles can be exploited by malicious actors.\footnote{This strategy of wielding multiple distinct identities to manipulate a system, especially when carried out on peer-to-peer networks, is known in computer security as a ``Sybil attack.'' For the foundational treatment, see \cite{druschel_revised_2002}.}

Defenders of online freedom of expression sometimes view deception as an unfortunate but necessary cost of preserving users’ privacy \cite{branscomb_anonymity_1995}. Indeed, the Internet’s pioneers saw anonymity as a fundamental pillar of privacy and freedom of expression \cite{kosseff_united_2022}. They built its architecture to let people participate in digital spaces without disclosing their real identities. The benefits from this approach are numerous, and worthy of steadfast protection—for example, anonymity allows people in oppressive regimes to express their opinions without fear of retribution. However, these benefits have come at some cost—one such cost is a lack of accountability for deceptive misuse.\footnote{For an account of how these norms evolved on Usenet, see \cite{yelvington_why_2006}. A common approach to undesired content sharing on Usenet was to apply social pressure \cite{bellovin_early_2019}.}

Although deceptive activity has significantly taxed the Internet,\footnote{It is difficult to precisely measure the tax imposed by malicious activity. One 2012 estimate of the cost to US firms and consumers of online spam alone was \$20 billion \cite{rao_economics_2012}.} the network remains largely usable for ordinary users. Defenses such as spam filters, IP blocklists, firewalls, and vigilant security analysts have helped to detect and mitigate deceptive attacks.\footnote{After being caught executing a particular type of attack, a malicious actor can try again. However, this usually involves additional work, like compromising more accounts or configuring a different proxy to disguise one’s IP address. If the actor does not change their tactics, they may be caught again by the same detection methods \cite{gao_detecting_2010}.} Furthermore, the attacks themselves have been resource-constrained—their reliance on human labor has kept their scale and quality in check.\footnote{For an overview of the types of labor involved in carrying out influence operations---and how AI may transform these---see \cite{goldstein_generative_2023}.}

The wide availability of increasingly capable AI\footnote{In this paper, when we refer to ``AI,'' we mean the \textit{combination} of available AI models that can be used together for different purposes. Generally, improvements in AI capabilities come from introducing new models that outperform existing ones in some way or offer better performance relative to their cost.} may upset this balance. Although bad actors have perpetrated deceptive attacks for decades, actors’ increased access to sophisticated and inexpensive AI tools may make their attacks far more effective---harder to distinguish and also more prevalent \cite{goldstein_generative_2023}.

The resulting escalation in deceptive activity across social media, public comment systems, and other essential digital services could create substantial challenges for institutions that rely on the Internet. Under pressure, these institutions may resort to invasive measures for verifying users’ identities online, overturning the Internet’s longstanding commitment to privacy and civil liberties.\footnote{China’s Ministry of Public Security and its Cyberspace Administration recently proposed a national system to associate online activity with individuals' legal identities \cite{tobin_china_2024}. The proposal has faced criticism from privacy advocates.}

In this paper, we focus on one potential solution to counter AI-powered deception: personhood credentials (hereafter referred to as PHCs), which certify that their holder is a person\footnote{For the purposes of this paper, we use ``person'' to refer to a human being, though we recognize that some domains use the term ``person'' more broadly. For example, in law, ``a person is any being whom the law regards as capable of rights or duties'' \cite{salmond_jurisprudence_1947}, which can include ``artificial persons'' (e.g., corporations) in addition to ``natural persons'' (i.e., human beings) \cite{garner_person_2024}. In moral philosophy, ``personhood'' need not apply only to human beings and could include other groups like non-human animals; ``personhood'' is sometimes associated with having full moral status, though philosophers disagree considerably about which traits are necessary or sufficient for full moral status \cite{clarke_rethinking_2021}. Though we use ``person'' and ``personhood'' to refer to human beings, this should not be read as implying any particular view on whether future AI systems should be considered moral persons or granted legal personhood (whether for moral or pragmatic reasons).} without revealing anything more about their identity. PHCs help to distinguish people from even the most advanced AI systems by relying on two important deficits of AI. Specifically, AI systems cannot convincingly mimic people offline, and they cannot bypass state-of-the-art cryptographic systems.\footnote{Cryptography relies on computationally hard mathematical problems, such as the factoring of very large numbers. There are not any known methods of efficiently solving certain such problems, whether by a human or an AI system.} PHCs allow people to interact with different digital service providers in a way that maintains their privacy through unlinkable pseudonymity—the credential holder can have a persistent, privacy-preserving pseudonym with different service providers, and different service providers have no way to link their activity. The issuer of the credential cannot trace their activity either.\footnote{While some national digital identity systems also draw on cryptography, they have very different goals and privacy affordances compared to PHCs. For example, Estonia's national ID-card allows citizens to conduct a wide range of activities online (e.g., cryptographically signing important contracts). But these systems are not intended to conceal \textit{which} holder of a card is performing the action \cite{noauthor_id-card_nodate}.} 

\begin{figure}[ht]
    \centerline{\includegraphics[width=.9\linewidth]{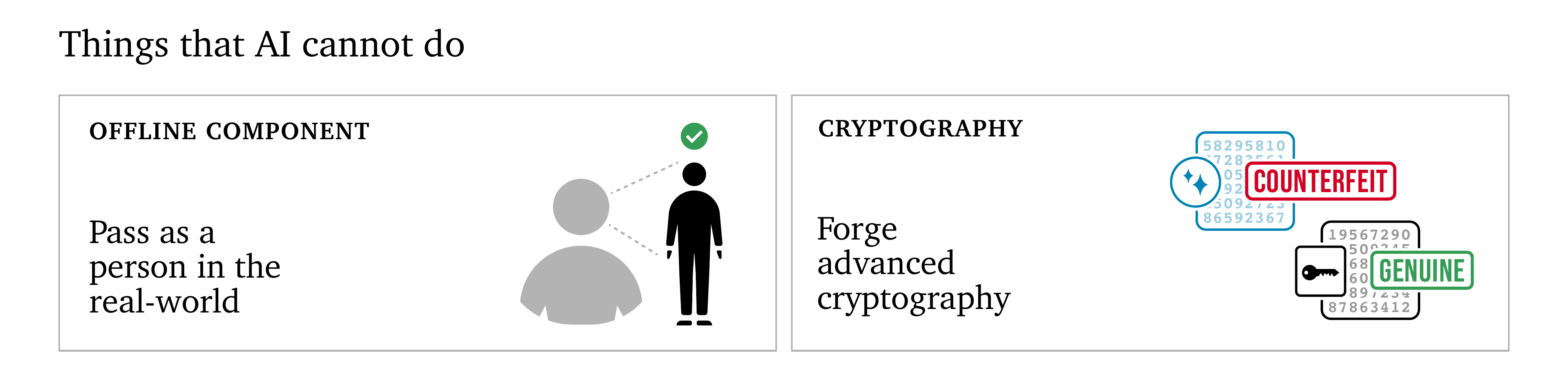}}
    \caption{Personhood credentials rely on two deficits of AI.}
    \label{fig:deception_protection}
\end{figure}

There are many possible issuers of such credentials, with many different methods an issuer can use to ensure that these credentials are available only to people and are not misappropriated by bad actors. For example, states could offer a PHC to each holder of a tax identification number, but thereafter have no way of tracing the uses of the credential.

That personhood credentials ought to exist is not a wholly novel idea. Related credentials are the subject of ongoing initiatives,\footnote{Our advocacy of personhood credentials aligns with many ongoing initiatives that aim for minimal disclosure identity and authentication systems, e.g., World Wide Web Consortium's (W3C's) Verifiable Credentials \cite{noauthor_verifiable_2024} and decentralized identifiers \cite{sporny_decentralized_2024},  European Union Digital Identity's (EUDI's) privacy-preserving digital wallets \cite{eudi_arf_2024}, and a range of standards and implementations of anonymous and attribute-based credentials \cite{noauthor_attributebased_nodate, hyperledger_anoncreds_2024,etsi}. One particularly relevant implementation is British Columbia’s Person Credential \cite{british_person_2024, bc_verfiable_2021}.} as well as decades of research in security and cryptography. In particular, a personhood credential is a type of anonymous credential---first proposed in the 1980s\footnote{An anonymous credential is a credential that allows its holder to prove some specific statement about themselves without revealing anything further \cite{chaum_untraceable_1981, chaum_security_1985, lysyanskaya_pseudonym_2000}. In the case of a personhood credential, the claim is that its holder is a person. In other cases, one could have an anonymous credential that proves their age without revealing anything further. Personhood credentials largely coincide with ``accountable pseudonyms,'' which aim to link a pseudonymous digital certificate that can be used anonymously online to an offline entity \cite{ford_offline_2008}. For a recent systematization of knowledge paper about anonymous credentials, see \cite{gunther_security_2023}.}---and can be understood as similar to a ``proof-of-personhood'' system.\footnote{The term ``proof of personhood'' emerged in blockchain communities---used in contrast to the concepts of ``proof of work'' and ``proof of stake''---as a strategy for dealing with Sybil attacks (where one entity subverts a digital system by using many pseudonymous identities to exert undue influence) in permissionless voting systems \cite{borge_proof_personhood_2017}. Given that ``proof of personhood systems'' often have specific aims (for instance, to allocate a cryptocurrency or voting rights in a decentralized autonomous organization) and different resulting design goals (e.g., they often aim for global uniqueness, sometimes via biometrics), we use the more generic term ``personhood credentials'' throughout the paper.} What is unprecedented is the scale and urgency of the problems these credentials could address, and the harms to trustworthy interactions\footnote{By ``trustworthy digital interactions,'' we mean interactions in which the parties’ reasonable expectations are achieved: for instance, a trustworthy online ecosystem is one in which service providers are confident that their services are being used as intended, and users can access digital services free from fears of abuse, attack, and other harms. We further unpack the foundations of this concept in \Cref{apx:conceptual}, drawing largely on \cite{jain_contextual_2024}.} on the Internet that may occur if unmitigated.

\subsection{Trends in AI threaten to make online deception more effective}
\label{ssec:deception_effective}

\begin{figure}[ht]
    \centerline{\includegraphics[width=.9\linewidth]{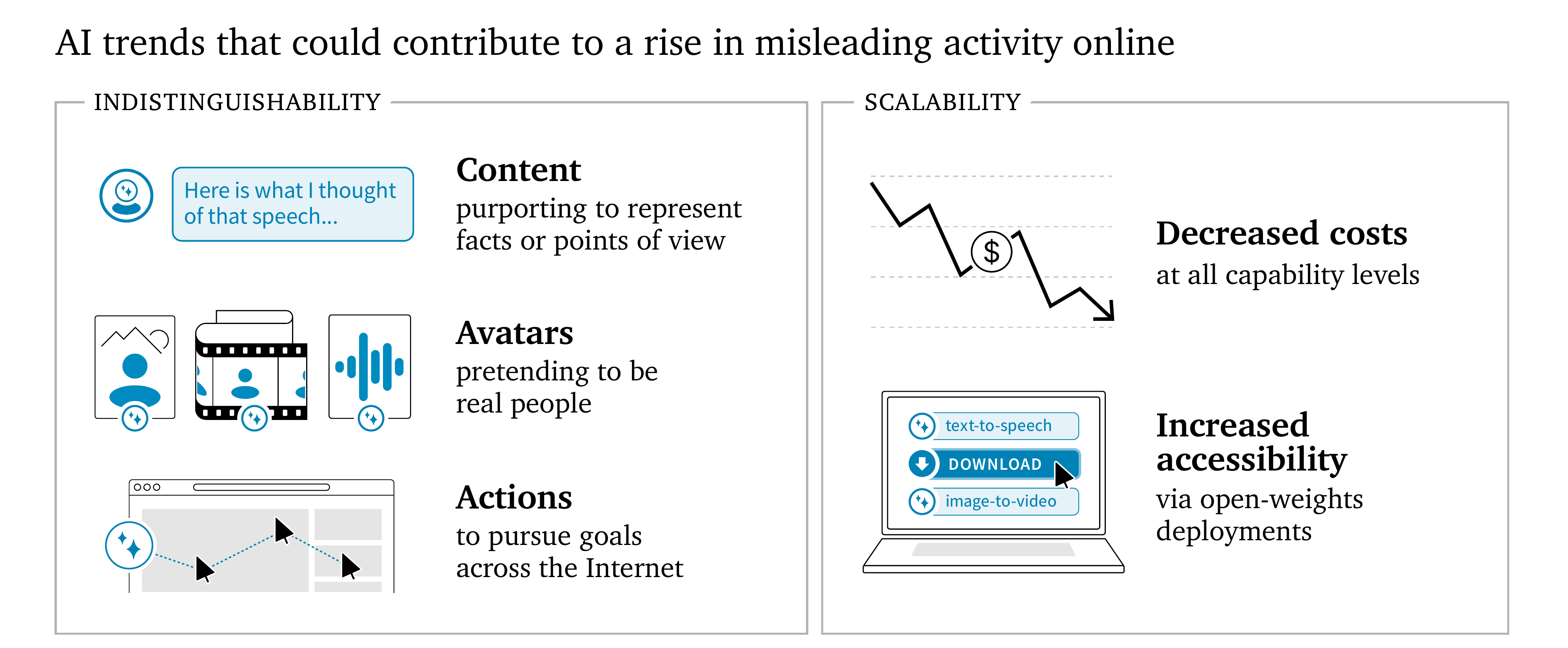}}
    \caption{Indistinguishability and scalability could drive an increase in AI-powered deception.}
    \label{fig:indistinguishability}
\end{figure}

To better understand the urgent need for personhood credentials, we discuss two notable trends in AI development that may contribute to more effective online deception.\footnote{When we refer to ``highly capable AI'' systems, we mean AI systems that exhibit these properties.} These trends are summarized in \Cref{fig:indistinguishability}.

First, users powered by AI systems are \textbf{increasingly indistinguishable} from people online. For example, AI-powered accounts can be populated with realistic human-like avatars that share high-quality content and that take increasingly autonomous actions.

Second, AI is becoming \textbf{increasingly scalable}---both more affordable and accessible, which can help for achieving many benefits but can also enable a larger amount of deception online.

\subsubsection*{\textit{AI systems are increasingly indistinguishable from people online}}
\label{sssec:indistinguishable}

Malicious actors can employ AI at many stages of their deceptive schemes, drawing on AI’s ability to seem like a real person online. This increasing indistinguishability comes from three categories of capabilities:\footnote{Malicious actors can draw on these capabilities at many stages of their deceptive schemes. For example, a malicious actor could create a pipeline of sockpuppet accounts on a social media site, in which it pairs a realistic-looking AI-generated profile picture (appearance) with a stream of posts that express a particular point-of-view (expressions) and a sequence of how to interact with organic users on the site (actions).}

\begin{itemize}[noitemsep,topsep=0pt]
    \item \underline{Generating human-like content} that expresses human-like experiences or points of view (e.g., posting commentary like ``Here is what I thought of that speech''). 
    \item \underline{Creating human-like avatars} through photos, videos, and audio (e.g., simulating a real-looking person on a video chat).
    \item \underline{Taking human-like actions} across the Internet (e.g., browsing websites like an ordinary user, making sophisticated plans to achieve goals they are given, solving CAPTCHAs when challenged).
\end{itemize}

AI systems are improving in each of these categories. For instance, AI-generated text and images are increasingly difficult to distinguish from human-created content in domains like politics,\footnote{For a study of legislators’ response rate to emails drafted by AI compared to those drafted by people—finding only a small difference—see \cite{kreps_potential_2023}.} art \cite{roose_aigenerated_2022}, and literature \cite{Clarke_scifi_2023}. Already, many people struggle to discern whether they are conversing by text with a human or an AI system in some settings.\footnote{Research suggests that text written by AI systems are hard to distinguish from text written by people, such as in Turing Test conversations \cite{jones_does_2024, noauthor_ai21_nodate}. These findings are not wholly new; at 2020’s GPT-3 launch, OpenAI reported that participants in a research study had trouble distinguishing between news articles by people versus those generated by AI \cite{Brown_fewshot_2020}.} A growing body of literature explores how AI-generated content might contribute to the efficacy of deceptive schemes, such as through AI's potential abilities to make persuasive arguments \cite{palmer_large_2023, bai_artificial_2023, anthropic_measuring_2024,elsayed_mechanism_2024}, tailor persuasion to a specific recipient \cite{matz_potential_2024}, and roleplay as certain personas \cite{wang_rolellm_2024}.

Likewise, AI is improving at creating human-like avatars. AI can enable malicious actors to create realistic high-quality photos and videos of people who do not exist \cite{korshunov_deepfakes_2018, bray_testing_2023, homeland_increasing_2021, brooks_video_2024}; animate photos of people into videos \cite{Siarohin_NeurIPS_2020}; take on the appearance of a different person over video chat \cite{xu_vasa_2024, horvitz_horizon_2022}; and speak with a realistic human voice, whether a generic voice or that of a specific person.\footnote{For an overview of voice generation abilities and resultant challenges, see \cite{openai_navigating_2024}. There are many different types of risks that can emerge with voice cloning \cite{hutiri_not_2024}. Voice cloning scams, in particular, are rampant and on the rise \cite{noauthor_aiaaic_nodate}.} These capabilities can be used to spoof identification checks, such as those requiring a selfie with a matching driver's license.\footnote{Some platform verification processes request proof of a government ID card—but even this process cannot restrict entry of AIs working on behalf of bad actors, as AI tools can be used to spoof a realistic-looking selfie of any person holding their supposed driver’s license \cite{cox_inside_2024}. If a service provider is willing to pay for more expensive verifications---such as those offered through the American Association of Motor Vehicle Administrators (AAMVA) in the US---they may be able to filter out some fraudulent IDs, but even these services have limits, such as not offering comparisons against the photo-of-record \cite{aamva_driver_nodate}.} In the near future, during a video chat, a user may not be able to discern whether their conversational partner is actually the person they are seeing, a person disguising themselves using AI, or even a complete AI simulation of a real or fictitious person.\footnote{There have already been incidents involving convincing deepfaked video calls of executives \cite{chen_finance_2024} and politicians \cite{guardian2022deepfake}.}

AI also continues to improve at taking human-like actions across the Internet. For instance, AI systems can solve CAPTCHA puzzles to gain access to Web services that are meant only for humans.\footnote{For an overview of AI’s ability to solve CAPTCHAs, as well as historical routes to bypass CAPTCHAs, see \Cref{sapx:captchas}.} More generally, AI systems are becoming increasingly agentic: capable of dynamically and independently carrying out actions toward goals over extended periods of time, without humans being in the loop or pre-specifying their actions or subgoals \cite{shavit_practices_2023, roose_personalized_2023, sebin_babyagi_2023, noauthor_what_2023-1}. In contrast to existing AI systems—which function mostly as content-generating machines—some of these agentic AIs will be more akin to Internet users: capable of navigating the Web interactively like humans do, but much faster. Some AI developers have predicted that AI agents will be a significant portion of future Internet users.\footnote{Upon releasing Meta’s latest AI model, CEO Mark Zuckerberg predicted a future of hundreds of millions or even billions of AI agents working online on behalf of small businesses \cite{Zuckerberg_llama_2023}.} At first, we anticipate that these AIs will be distinguishable by anomaly detection systems that measure atypical Internet behavior on a website. But, over time, AI agents might be able to convincingly mimic the behavioral signals of real human users.\footnote{There are already instances of AI systems learning to imitate human behavior in complex digital environments \cite{baker_learning_2022}.} With greater dynamism of AI agents will come a larger surface area of risks: For instance, agentic AIs might be able to pursue many dangerous ``long con'' schemes at once—like creating a number of fraudulent personas to infiltrate open-source communities that manage vital digital infrastructure.\footnote{While this specific scenario is speculative, researchers are evaluating frontier models for signs of dangerous capabilities that could enable such long-con cyberattacks \cite{phuong_evaluating_2024, fang_agents_2024, fang_exploit_2024}. Social engineering attacks by humans---when they happen over a relatively long timescale---can be very resource-intensive \cite{kaspersky_social_2024, vijayan_attacker_2024}. AI agents may have significant advantages over humans in terms of their levels of patience and ability to multitask across many attacks at once.}

\subsubsection*{\textit{AI systems are increasingly scalable}}
\label{sssec:ai_scalable}

 AI systems are not only more capable but also increasingly scalable. Malicious actors can leverage AI tools to execute widespread attacks that they may not have had resources to execute previously. The same factors that lower barriers to beneficial access\footnote{Some beneficial uses of AI include helping people regain abilities that they have lost, such as using AI to power a voice for someone who has lost the ability to speak fluently \cite{sprunt_neurological_2024}. Managing AI risks can be challenging because the same AI applications can be either beneficial or harmful, depending on the scale at which they are used and the intentions behind them.} can also enable attacks to increase in scope and frequency.

One factor driving the scalability of AI-powered attacks is cost. AI models are becoming more affordable at every capability level.\footnote{AI systems’ abilities have significantly improved in recent years, with a clear relationship between investment and a system’s ultimate capabilities, often referred to as ``scaling laws'' \cite{kaplan_scaling_2020, hoffmann_training_2022, anderljung_frontier_2023, zhao_survey_2023}. At the same time that AI systems’ abilities have improved, the cost for a given level of ability has generally decreased. See \cite{nguyen_cost_2024} for a visualization illustrating how even relatively cheap AI models from 2024 outperform leading models from 2022.} This implies that malicious actors can generate content for a host of deceptive aliases at much lower cost. Thus, with reduced investment required for a successful attack, the frequency of attacks might increase.\footnote{For a framework of different factors that affect AI’s impact on influence operations, see \cite{goldstein_generative_2023}.} This dynamic is particularly relevant for operations that succeed through persistent and repeated efforts.

Another factor contributing to scalability is the ease of access: The release of highly capable open-weights models makes AI capabilities more available to both well-intentioned and malicious actors. Many open-weights models are available through user-friendly interfaces, decreasing the technical skill required for such uses.\footnote{For instance, some open-weights models are available through desktop applications that can be run on consumer hardware \cite{nomic_announcing_nodate}.} Compared to closed-weights counterparts, open-weights models may increase the relative ease of misuse because they offer less moderation and monitoring of relevant capabilities.\footnote{For a recent position paper on the risks and opportunities of open-source generative AI, see \cite{eiras_near_2024}. In the context of AI, open-source is often used to indicate releasing the weights and inference code for a model, but not necessarily the full source code by which the model was trained \cite{seger_opensourcing_2023}. For this reason, these models are sometimes called ``open-weights'' rather than open-source.} 

\subsection{Current solutions for countering AI-powered deception need improvement}
\label{ssec:need_improvement}

There are many tools currently used to reduce deceptive and malicious activity online, particularly when the activity is AI-powered. Here, we discuss how the addition of personhood credentials to the toolkit could improve on some tools and complement others, significantly bolstering the foundations of trustworthy interaction. We summarize the discussion in \Cref{table:existing_tools}.

\begin{table}[ht]
\centering
    \centerline{\includegraphics[width=.9\linewidth]{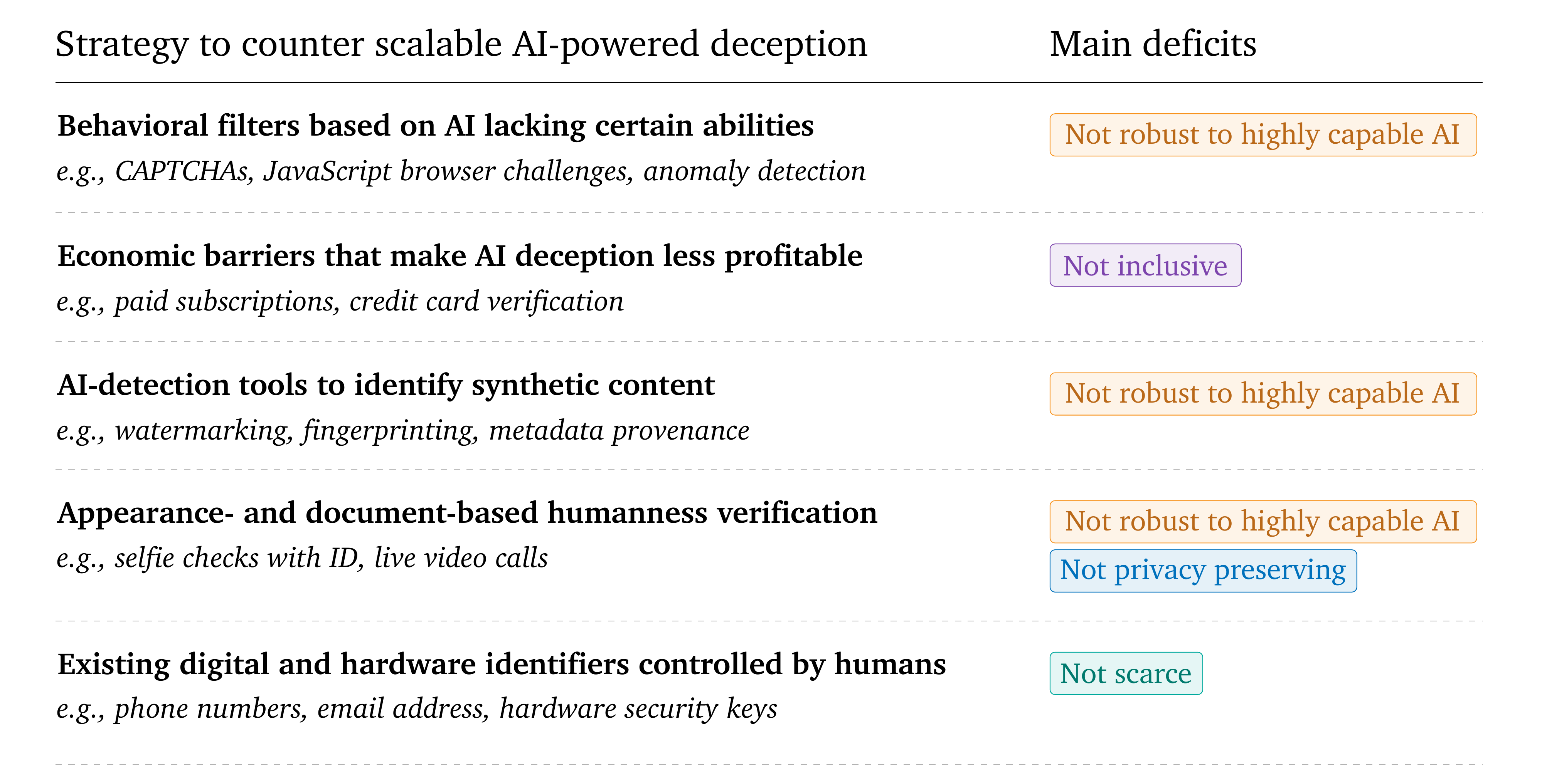}}
    \caption{Existing tools for countering AI-powered deception and their main deficits.}
    \label{table:existing_tools}
\end{table}

One approach is \textbf{behavioral filters}: distinguishing AI-powered activity based on human abilities or behaviors that are difficult for AI to imitate. For instance, CAPTCHAs historically exploited bots’ inability to consistently recognize distorted letters \cite{ahn_recaptcha_2008}. JavaScript-based browser challenges and anomaly detection systems lean on the fact that—again, historically—bots did not use Web browsers in ways that people do.\footnote{For instance, the patterns of fraudulent account registrations often differ from those of legitimate users \cite{yuan_detecting_2019}. For a survey of different browser challenges, see \cite{aminazad_webrunner_2020}.} These filtering methods are becoming less effective as AI systems improve; they are not robust solutions on their own.\footnote{In \Cref{sapx:captchas}, we offer a more detailed survey of behavioral filters and their limitations.}

A second approach is to impose \textbf{economic barriers} that make it costly to perpetuate AI-powered deception at scale. Some websites introduce paid subscriptions with the explicit intent of distinguishing between people and bots by making it more expensive to use many bot accounts.\footnote{Elon Musk reported this as one factor in his decision to institute a paid subscription option on Twitter (now X) \cite{sankaran_elon_2023}.} These payments, however, disproportionately affect lower-income users and are an insufficient barrier against particularly profitable forms of malicious activity \cite{germain_meta_2023, conversation_social_2023}. Relatedly, some digital services check whether users hold a valid credit card---without necessarily charging a payment to the card---though this approach can be circumvented and also has similar challenges with inclusivity.\footnote{Credit card requirements can be circumvented via virtual cards \cite{privacy_whatare_2024}.}

A third approach consists of tools that aim to \textbf{detect AI-generated \textit{content}}. Policymakers and technologists have directed investment toward developing ``synthetic content transparency'' methods in recent years.\footnote{For an overview of these methods, see \cite{pai_staff_building_2023}. These techniques were highlighted in President Biden’s Executive Order on the Safe, Secure, and Trustworthy Development and Use of Artificial Intelligence \cite{Executive-Order2023}, which resulted in a report from the National Institute of Standards and Technology (NIST) that provides more detail on these techniques \cite{barker_reducing_2020}. In \Cref{sapx:synthetic_transparency} we offer a fuller discussion of synthetic content transparency techniques, comparing their affordances to those of PHCs.} Watermarking, for instance, embeds signals in AI-generated digital content; these signals can help to identify whether content is the output of an AI model. Such tools are useful in many circumstances, but they also have shortcomings: For instance, adversaries may alter AI outputs to try to evade detection, and the tools are not perfectly accurate even if content is undisturbed by an adversary \cite{zhang_watermarks_2023, kelly_watermarks_2023}.\footnote{In some open-source model implementations, the watermarking function can merely be removed from the code before running \cite{noauthor_stable_2023}.}  More generally, focusing only on whether content is AI-generated does not address other aspects of trustworthy interaction: Consider an army of AI-powered sockpuppet accounts that amplify human-generated content, initially posted from real people’s accounts, to push a specific agenda. There is no AI-generated content present---only unobservable decisions about which human-created content to repost---and yet the activity is deceptive.

A fourth approach is to use \textbf{document- or appearance-based verification} to confirm that there is a person behind some digital activity. Some verification protocols require that users join a video chat or show evidence that they are in possession of physical identifying documents (for example, they may ask users to send a selfie holding a matching driver’s license \cite{stobierski_what_nodate}). At times, these methods go too far toward identifying ``who'' is behind some activity—collecting more information than required to merely verify that there is ``a'' person conducting the digital activity. Such approaches are not only potentially intrusive from the user’s perspective, but also involve the collection of sensitive personal identifying information, which can introduce important security concerns. Beyond privacy and security issues for users and service providers, such approaches are also not robust to newer AI systems, which are increasingly capable of creating content and avatars that pass these checkpoints.\footnote{For an example of a particular case in which even a video call did not successfully head off AI-powered deception, see \cite{olson_zoom_2024}. See also \cite{cox_inside_2024, lanz_people_2024}. Beyond using AI to spoof such checks, malicious actors can also purchase photos of people holding up their passports for validation \cite{quito_theres_2018}.}

Finally, digital service providers often use \textbf{identifiers} like emails, phone numbers, and purpose-built \textbf{hardware-based authenticators} to try to verify that there is a person taking actions on their services. For example, one-time passwords sent via SMS can offer some indication that the entity behind some digital activity has access to a physical device---though given the rise of tools for acquiring virtual phone numbers and handling complex workflows through these phone numbers, this indication is not particularly strong.\footnote{For instance, Twilio, a company that provides cloud communications services, has been used to acquire phone numbers used in scams \cite{anders_fcc_2023}.} There is a more general problem at play: requiring unique phone numbers, email addresses, and even hardware authenticators at signup can reduce some duplicate account creation, but none of these identifiers are scarce enough to establish that their holder is distinct from other users (and thus that the holder could not be creating deceptive identities at scale).\footnote{For more detail on creating fraudulent accounts that require phone number validation, see \cite{thomas_dialing_2014}. Requiring such validation can increase the cost of deception but is not a sufficient solution.} Moreover, some of these identifiers facilitate digital tracking \cite{chen_everyone_2023} and are less private than we aim for with PHCs.

Thus far, we have described trends in AI that could contribute to scalable deception online, as well as reasons why current solutions may be insufficient for countering it. In the remainder of the paper, we articulate a case for personhood credentials as a foundational tool in the toolkit for protecting a trustworthy digital ecosystem, even as AI grows more advanced.

\section{Defining personhood credentials}
\label{sec:definition}

In this section, we outline the design requirements of a personhood credentialing system. Then, we discuss how these requirements balance inherent tensions between preserving user privacy and reducing the possibility of deceptive activity at scale. In \Cref{apx:implementation}, we briefly discuss a number of potential implementation approaches---we leave fuller evaluation of such approaches for further research.\footnote{We discuss a range of specific ways to implement credential issuance and usage in \Cref{apx:implementation}. One possible implementation option is to build personhood credentials atop the W3C global standards for Verifiable Credentials (VCs), which are a means of expressing a variety of traditional credentials—e.g., driver’s licenses, university degrees, passports—on the Web, in a way that is ``cryptographically secure, privacy respecting, and machine-verifiable'' \cite{noauthor_verifiable_2024}.}

\begin{figure}[ht]
    \centerline{\includegraphics[width=1\linewidth]{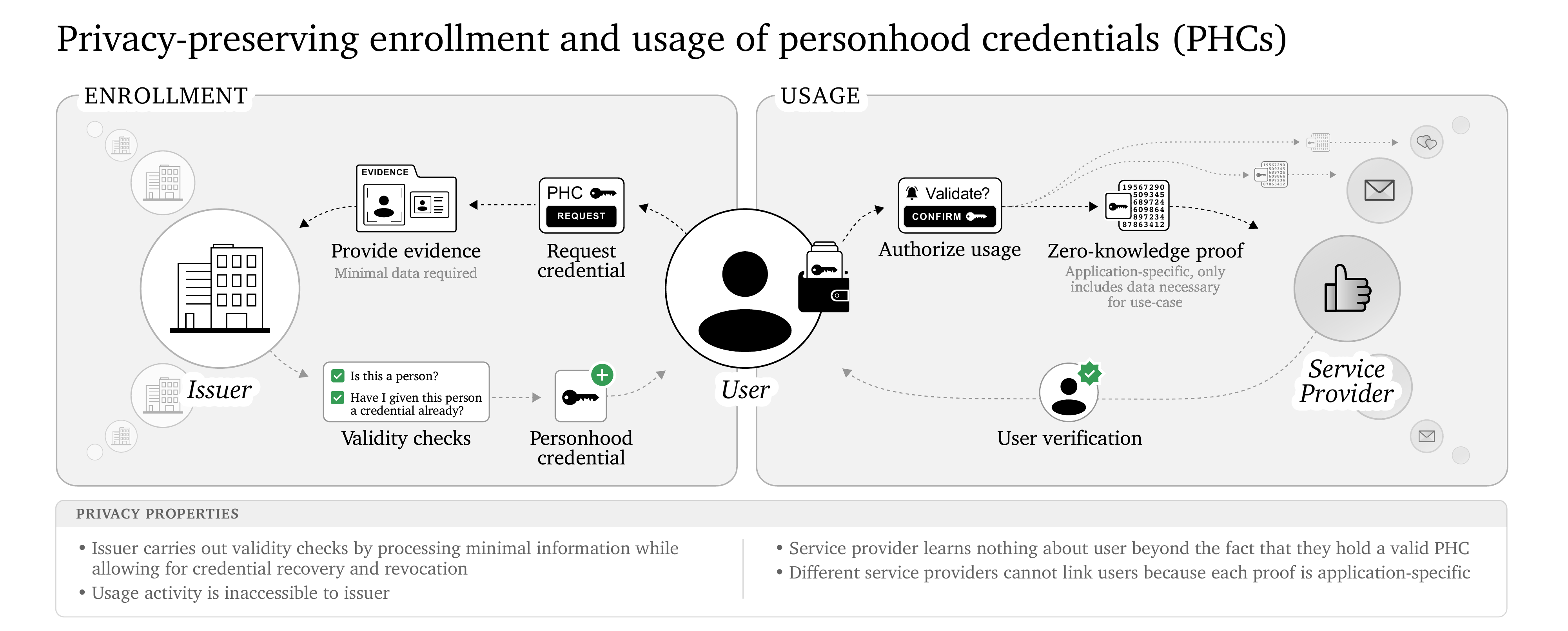}}
    \caption{Illustration of enrollment and usage of a personhood credential. }
    \begin{center}
    \vspace*{-.4cm}
    \begin{minipage}{.8\textwidth}
       \small{\textbf{Note:} There are two stages of a personhood credential system.  First, the user enrolls. Second, the user can then use the credential with a range of service providers without needing to re-enroll.}
    \end{minipage}
    \end{center}
    \label{fig:enrollment_usage}
\end{figure}

A personhood credential digitally certifies that an issuing entity (\textbf{``issuer''}) believes its holder to be a real person who has not received a credential from them previously. The issuer issues a credential through a process we refer to as ``enrollment.'' A third-party digital service (\textbf{``service provider''})\footnote{We use ``service provider'' to mean ``relying party,'' the more common term in security communities.} can request evidence that a user holds a personhood credential as part of some authorization process (like receiving up to a certain number of verified accounts); we refer to this process as ``usage.'' The enrollment and usage processes are illustrated in \Cref{fig:enrollment_usage}.

When a holder uses their personhood credential, they prove to the service provider that they hold a valid credential without revealing the credential itself (e.g., through a zero-knowledge proof \cite{goldwasser_knowledge_1989}). When necessary, the holder also has the ability to prove that the credential has not yet been used for this particular service (or has been used fewer times than the service's limit). In either case, using a PHC does not reveal any aspect of its holder’s legal identity.\footnote{PHCs are a type of digital credential, which we define to be an electronic assertion made by an issuer and intended to be used more than once. This definition aligns closely with NIST's definition of digital credential \cite{noauthor_verifiable_2024}. PHCs could optionally be used in combination with other forms of credentials---like attestations of where one went to college, or where one currently lives---for use-cases that rely on proving further aspects about oneself.}

\subsection{Foundational requirements of a personhood credentialing system}
\label{ssec:foundational_requirements}

A PHC system meets the following requirements:

\begin{enumerate}[noitemsep,topsep=0pt]
    \item \textbf{Credential limits (1 credential per person per issuer)}: The PHC issuer aims to issue only one credential per person and provides ways to mitigate the impact of transfer or theft of credentials.
    \begin{enumerate}[a.,noitemsep,topsep=0pt]
        \item \underline{Issuers check one-per-person requirement at enrollment}: The issuer has an effective check of whether a person has already received a personhood credential from them.\footnote{Often this check will include some form of  evidence that depends on an interaction that occurred in real life, so that AI systems cannot be dispatched to obtain credentials. This evidence \textit{need not} involve a direct in-person interaction between the user and the issuer at the time of enrollment. For example, a passport could be used in an enrollment process, providing evidence that its holder at some point went through the necessary in-person steps to acquire such a document. We offer further discussion of methods for satisfying the credential limit in \Cref{sapx:methods_limit}.}
        \item \underline{Expiry or regular re-authentication}: To mitigate the theft or transfer of credentials, there is a periodic process designed to reduce credential use by someone other than the original holder.\footnote{This regular re-authentication could be achieved through a combination of factors, such as the continued possession of some root document (e.g., a passport), or through tight expiration limits. }
    \end{enumerate}

\vspace{2mm}
    \item \textbf{Unlinkable pseudonymity (privacy)}: PHCs let a user interact with services anonymously through a service-specific pseudonym; the user’s digital activity is untraceable by the issuer and unlinkable across service providers, even if service providers and issuers collude.\footnote{We discuss some cryptographic methods that could be used to achieve the requirements outlined here in \Cref{sapx:methods_unlinkable}.}
    \begin{enumerate}[a.,noitemsep,topsep=0pt]
        \item \underline{Minimal identifying information stored during enrollment}: The issuer associates minimum necessary identifying information\footnote{The minimum necessary amount of information will vary depending on the types of recovery and revocation procedures the system wishes to afford. Setting tight expiry limits is one way of minimizing the information needed for revocation---see related discussions in a recent comment \cite{baum_cryptographer_nodate} on the European Digital Identity Wallet Architecture and Framework \cite{eudi_arf_2024}, and in standard ISO/IEC 18013-5 on mobile driving licenses \cite{noauthor_personal_2021}.} between a specific personhood credential and its holder.
        \item \underline{Minimal disclosure during usage}: When a user presents evidence of a personhood credential to a service provider, it reveals to the service provider nothing more than ``this person holds a valid PHC'' or, with the user’s authorization, ``this person holds a valid PHC not yet used with this service.''
        \item \underline{Unlinkability by default}:\footnote{By unlinkability, we mean unlinkability via direct use of one’s personhood credential. Today, users are sometimes identifiable across websites through factors like their Web browser, even without sharing other identifiers \cite{eckersley_how_2010}. PHCs are not a direct solution for commonplace tracking of users across the Internet or aggregation of this information via data brokers. However, to the extent that such tracking is facilitated by reusing identifiers across sites (like email addresses or phone numbers), PHCs might make it more difficult to create linkages than previously.} By default, service providers or issuers cannot trace or link usage activity across uses, even if issuers and service providers collude.\footnote{One risk of government-issued personhood credentials is that the government may be able to compel service providers to turn over records, even if a service provider does not wish to. To be fully privacy-preserving, a personhood credential must be resistant to these forms of subversion.} The issuer, by default, learns nothing when a PHC has been used. Service providers do not learn anything when a PHC that has been used with their service is used with another service.
    \end{enumerate}
\end{enumerate}

In \Cref{apx:implementation}, we elaborate on the assumptions and infrastructure needed to implement a system that meets these requirements.

\subsection{Credential limits and the goals of a personhood credentialing ecosystem}
\label{ssec:credential_limits}

A personhood credentialing ecosystem has two core goals, which are sometimes in tension: to reduce scaled deception while also protecting user privacy and civil liberties.

Maintaining a one-per-person per-issuer credential limit helps to balance the ecosystem’s goals: attaining reasonable limits on the scale of deceptive activity any person can carry out, while preserving a meaningful degree of choice for users. 

We argue that this balance is achieved when there are multiple issuers, each of which limits the number of credentials a person can receive from them. Each person can then obtain a bounded number of credentials—more than one, to counter risks to privacy and civil liberties, but not so many that the credential loses its ability to prevent scaled deception.

\begin{figure}[ht]\centerline{\includegraphics[width=.9\linewidth]{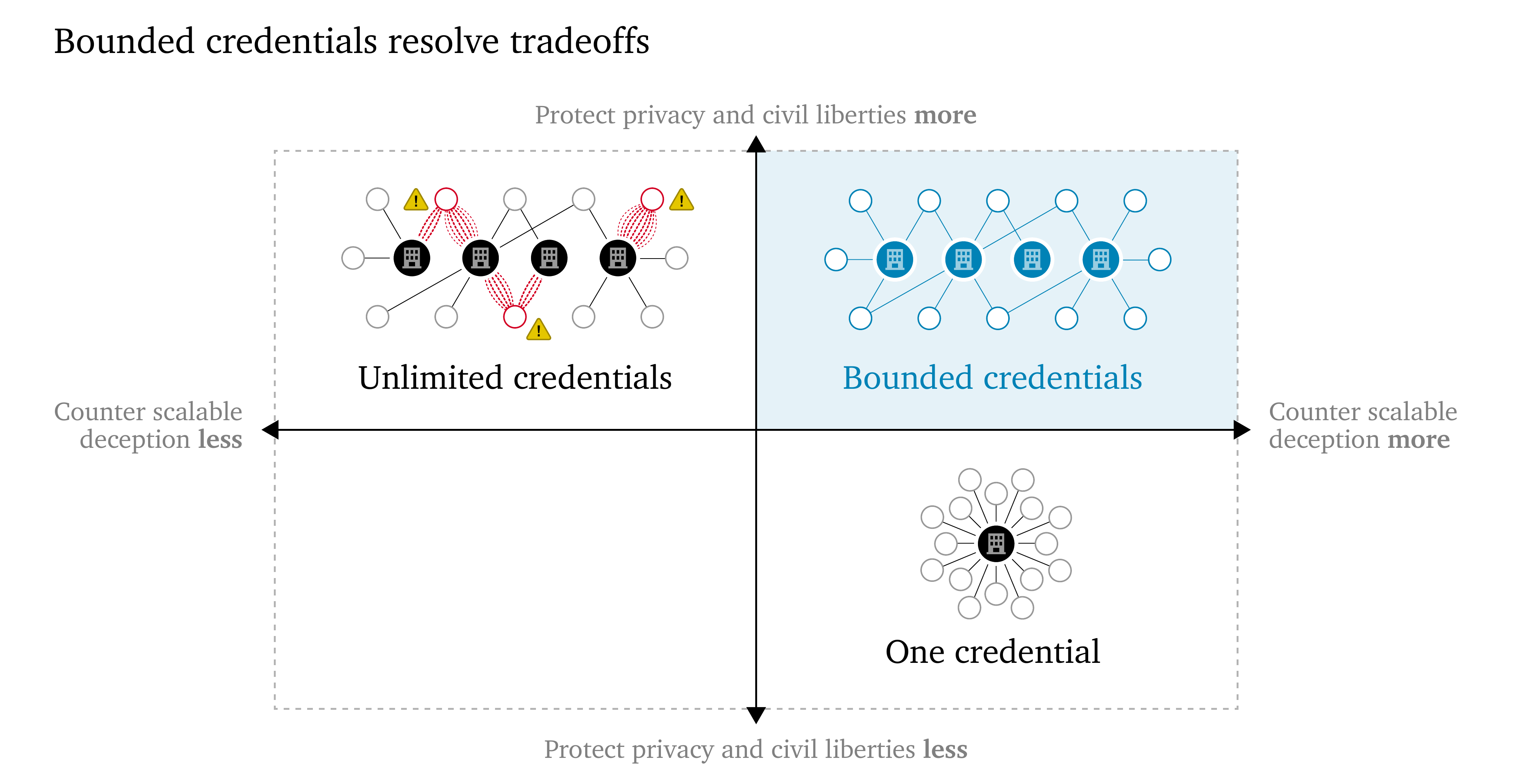}}
    \caption{Ecosystem design trade-offs---the argument for bounded credentials.}
    \label{fig:bounded_credentials}
    \begin{center}
    \vspace*{-.4cm}
    \begin{minipage}{.8\textwidth}
       \small{\textbf{Note:} Open circles represent individuals, and filled circles represent issuers. Top left: Unlimited credentials per person (red). Less effective against deception but better for privacy and civil liberties due to minimal issuer data storage. Bottom right: One credential per person, from one issuer. Effective against deception but risky for privacy and civil liberties. Top right: One credential per person per issuer, with multiple issuers. This balances privacy protection and countering deception. }
    \end{minipage}
    \end{center}
\end{figure}
To illustrate this, we further deconstruct each ecosystem goal into two factors that an ecosystem should try to attain:

\textbf{Reduce the harms from scaled deception.}
Limit the scale at which any actor can deceive others, via:

\emph{Per-person rate limits}. Without per-person limits on some digital activities, harmful actors can pose as multiple people to gain outsized influence on peer-to-peer platforms and in public comment processes.\footnote{For example, the public comment process for the FCC’s 2017 decision on repealing net neutrality protections was compromised by nearly 18 million bot-generated comments \cite{james_report_2021}.} In addition, harmful actors can evade a service’s rules---by, for example, creating many new accounts even when caught violating a service’s policies.\footnote{The market size of the underground ecosystem for account registration, by one estimate, is roughly 4.8--128.1 million USD per year (as of 2022) \cite{gao_demystify_2022}. Illicit account creation and the repeat abuse it enables are taxes on digital services and their legitimate users.}

\emph{Limited incentives for credential transfer.} When there are large incentives for theft or sale of credentials, a credential’s signal of real personhood erodes over time. Likewise, if it is possible to clandestinely lend one’s credential, ``authentication farms'' may emerge that help bots bypass personhood checks---eventually the credential is not a persistent signal of personhood at all.\footnote{If personhood credentials can be used with a service without limits (e.g., used to obtain many verified accounts), we expect authentication farms to emerge, in which people verify their PHC with a service and then hand off their account to an AI-powered user. This is comparable to how today’s bots can route CAPTCHAs to humans who unblock the bots’ path for a fraction of a cent. We discuss these practices in \Cref{sapx:captchas}.}

\textbf{Preserve privacy and civil liberties.}
Provide users a meaningful choice of issuers and system features that guard their privacy, via:

\emph{Minimal information processing and storage.} Limiting the amount of personal data processed and stored by credential issuers is essential to preserving user privacy. By minimizing data collection and retention, the ecosystem reduces the risk of misuse or unauthorized exposure of sensitive information.

\emph{Checks on power.} Preventing the concentration of authority over digital credentials is essential for safeguarding civil liberties, so that no entity can unilaterally dictate terms or exploit personal information stored or processed in the credentialing system.

In \Cref{apx:ecosystem}, we highlight some inherent tensions between these ecosystem design goals through two extremes, which we disfavor: one in which people can acquire an unlimited number of credentials, and one in which people can acquire only one credential from a single issuer. Through these extremes, we illustrate that an ecosystem with bounded credentials may best resolve these tensions. \Cref{fig:bounded_credentials} summarizes the argument.

\section{Prospective benefits of personhood credentials}
\label{sec:benefits}

\begin{figure}[ht]
    \centerline{\includegraphics[width=.9\linewidth]{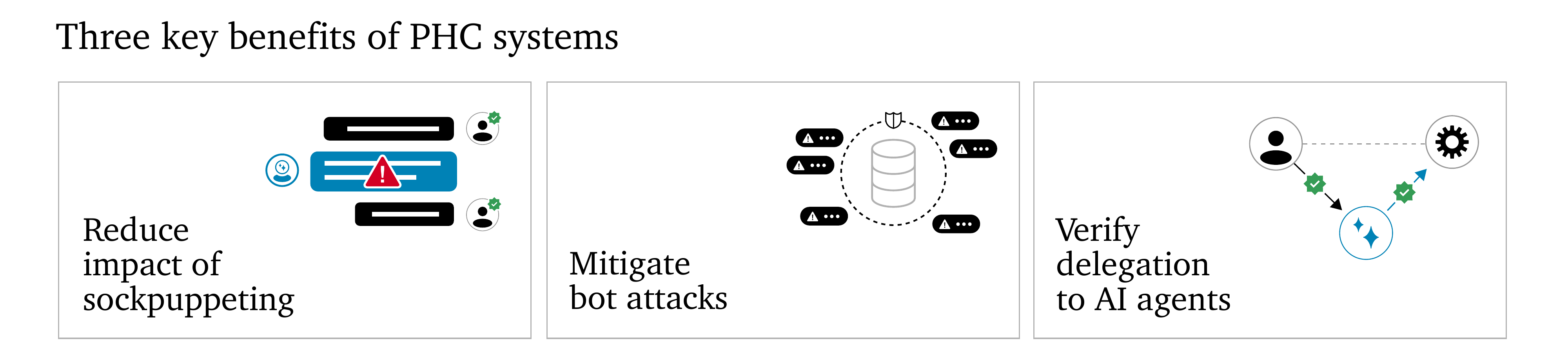}}
    \caption{Summary of three key benefits of PHC systems.}
    \label{fig:key_benefits}
\end{figure}

We highlight three ways that well-designed PHC systems can reduce the impact of scalable deception online. These benefits are summarized in \Cref{fig:key_benefits}. PHCs can help to:

\begin{itemize}[noitemsep]
    \item \underline{Reduce the impact of sockpuppeting}:
    \hspace{\parindent} Enable authentic input and engagement from real people at scale, free from deceptive profiles representing people that do not actually exist.
    
    \item \underline{Mitigate bot attacks}:
    \hspace{\parindent} Prevent coordinated attacks of bots circumventing platforms’ rules to continue abusing Web services.

    \item \underline{Verify delegation to AI assistants}:
    \hspace{\parindent} Signal that an AI assistant is a delegate of a trustworthy person, as opposed to a delegate of a malicious actor.
\end{itemize}

\subsection{Reduce the impact of sockpuppeting}
\label{ssec:reduce_sockpuppeting}

Malicious actors lean on ``sockpuppetry''---adopting the persona of some hypothetical ``person'' who could exist but does not—for a range of purposes:\footnote{Note that many AI-powered personas are benign and beneficial, particularly when transparently disclosed. For instance, an AI-powered chatbot may transparently mimic a science tutor to teach students more effectively. This illustrates a challenging aspect of detecting deceptive uses: An educational institution deploying an AI-powered science tutor is similar in many regards to a bad actor deploying an AI-powered scientist to spread medical falsehoods.} manipulating perception of public political opinion on social media, propping up (or attacking) the reputation of businesses or individuals, and carrying out scams on digital marketplaces \cite{diresta_tactics_2019, web_understanding_2021, keller2020political,Zhang2013OnlineAA}. Historically, sockpuppets have been powered either by basic automation or by people in low-wage countries paid by malicious actors to control many misrepresented personas.\footnote{For one report of deceptive accounts in West Africa purporting to be US-based to influence the US political climate, see \cite{ward_russian_2020}.}

Personhood credentials are a mechanism for reducing the viability and impact of sockpuppets. Regardless of whether sockpuppets are AI-powered, services that adopt PHCs can reduce malicious actors’ ability to falsely present themselves as multiple individuals on their platform by, for example, imposing account verification limits on each personhood credential.\footnote{An account verification limit does not strictly need to be ``one verified account per PHC'' in order for PHCs to curb malicious activity: A website could decide, for instance, that it wishes for each holder to have at most three verified accounts, which could be unlinked to one another. The important property here is the ability to enforce some finite limit by counting how many times a PHC has been used for a purpose, without more specifically identifying the holder or their particular PHC.} Once a user is verified to be a person, there are many ways a site could incorporate PHCs to facilitate trustworthy discourse, e.g., by boosting or labeling such accounts, or by offering users ways to screen out accounts that are not verified.\footnote{Stronger distinguishability could facilitate a more trusting online community. On today’s Internet, sometimes false accusations are made that a given account is a bot \cite{assenmacher_you_2024}. Knowing another user to in fact be a person could inspire greater trust and ultimately lead to more productive conversations. For instance, some research indicates that using a profile picture with a face in it leads to a greater empathetic response from others in online discussions \cite{liu_how_2022}. PHCs may help achieve a similar effect.} \Cref{scenario:reduce_sockpuppet} outlines one hypothetical use. Such methods are not meant to entirely rid a service of AI-generated content or AI-powered accounts, which can often be benign.\footnote{Some services have established processes for the appropriate disclosure of automated accounts, such as accounts operated with AI or through other computerized systems that do not require human intervention \cite{xhelp_about_nodate}.} But PHCs do offer a range of options for curbing the impact when AI’s use is deceptive---particularly at scale.\footnote{While the introduction of PHCs does not entirely solve the problem of AI-powered deception, it dramatically reduces the scale. For example, someone might still hand over their single PHC-verified dating site account to an agentic AI, which they direct to carry out a catfishing scam. Yet the catfisher is less likely to be successful with a single profile as opposed to many potentially targeted profiles used in parallel. Or, someone may post misleading AI-generated content from their single PHC-verified social media account to spread disinformation. But the spreader of disinformation will not be as successful in making their misleading content go viral without amplification from other independent-seeming accounts that they also control. In both cases, if the bad actor is found to be in violation of the service’s rules, they can be prevented from returning---a dynamic we discuss further in the following subsection.}

\begin{scenario}
    \centerline{\includegraphics[width=.9\linewidth]{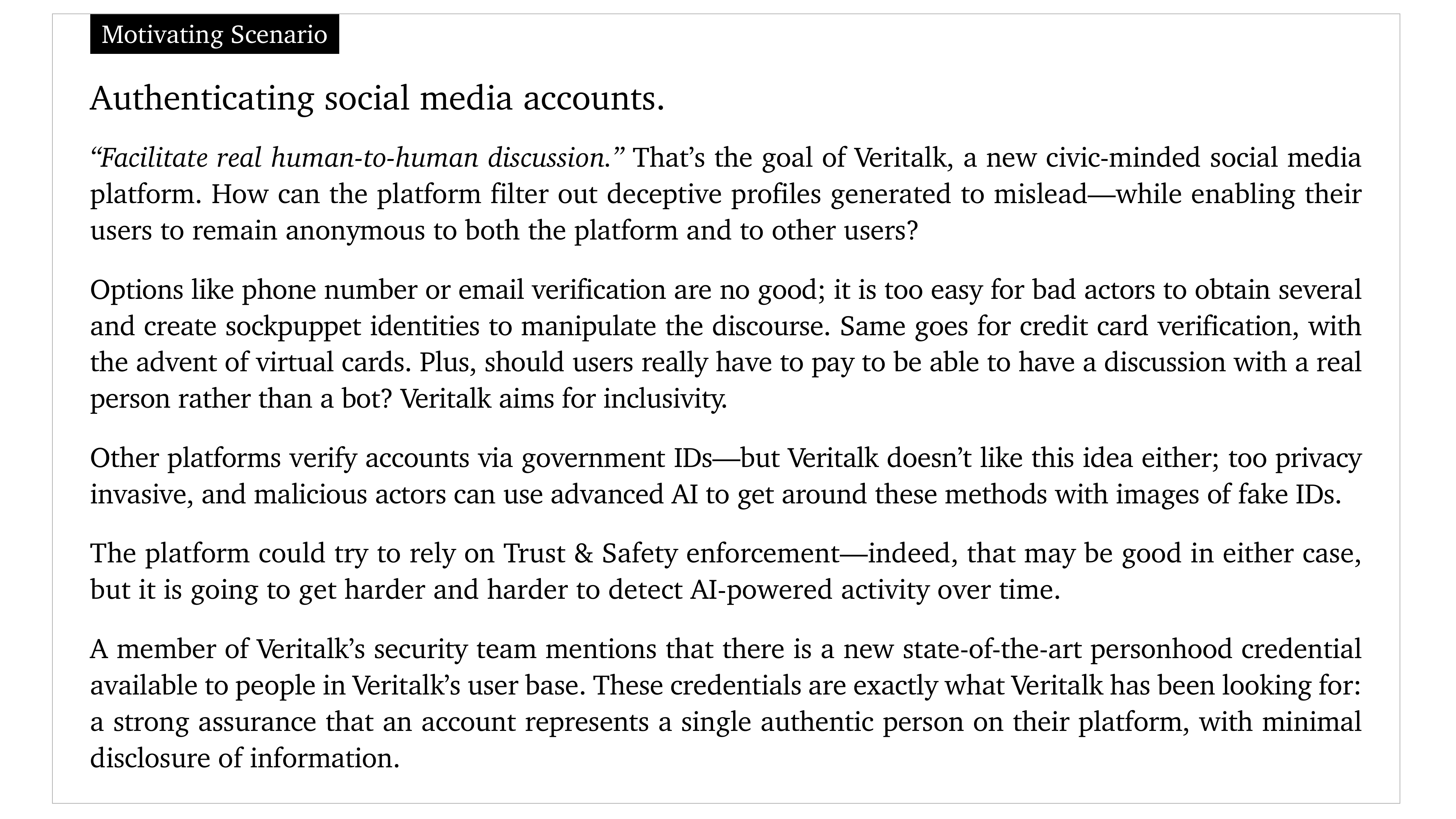}}
    \caption{Authenticating social media accounts.}
    \label{scenario:reduce_sockpuppet}
\end{scenario}

In recent years, sockpuppets have caused harm across a number of settings. On social media, sockpuppets attempt to manipulate public opinion on a topic \cite{hagen2022rise, mckay2021disinformation, bastos_public_2018,marlow2021bots}.\footnote{For a broader investigation of sockpuppets in online communities, see \cite{kumar_army_2017}.} \footnote{Some digital services describe the rationale for their sites' rules in terms of reducing the impact of sockpuppets. X’s Community Notes, for instance, states a requirement for accounts to have a unique phone number to ``help prevent the creation of large numbers of fake or sock puppet contributor accounts that could be used for inauthentic rating'' \cite{x_community_nodate}.} In online dating, bot-powered profiles are used for catfishing, luring users into costly and sometimes dangerous situations \cite{coffey_am_2024}. In formal government comment processes, bots overwhelm genuine civic input.\footnote{For example, millions of fake comments were submitted to the Federal Communications Commission’s 2017 public request for comment on net neutrality. Researchers have conducted analysis on the methods used to generate such inauthentic comments, which were far more rudimentary than is possible with today’s AI \cite{weiss_investigation_2020}. Sockpuppets continue to overwhelm US government public comment processes, and few agencies have taken action to systematically filter out or catch the malicious actors perpetrating these abuses \cite{senate_abuse_2019, panditharatne_artificial_2023}. Recently, a bill aiming to curb foreign and AI-powered abuse in the public comment process for the Bureau of Land Management (the American Voices in Federal Lands Act) was introduced in the US Senate \cite{barrasso_bill_2024}.}

By reducing the impact of sockpuppets, PHCs might also help to advance a wide range of beneficial experiments in digital governance in the public, private, and nonprofit sectors, though these are speculative and also come with risks.\footnote{Scholars have argued that new democratic innovations may be especially important in the age of highly capable AI \cite{allen_real_2024, bernholz_digital_2021}.} For instance, AI tools might help to digitally gather and analyze public opinions from large segments of the population \cite{fish_generative_2023}---PHCs can help ensure that people do not use multiple personas to wield more influence and thereby can increase the perceived legitimacy of such processes.\footnote{PHCs can help digital governance processes be more robust to claims that AI bots were used to influence a particular outcome. For more on the ability to cast doubt on outcomes by invoking the specter of AI manipulation, see \cite{schiff_liars_2022}.} In particular, PHCs can reduce the impact of communities that might otherwise intentionally undermine these processes.\footnote{One challenge in soliciting broad input online is that particular communities might flood a poll to try to undermine the poll’s intent. For instance, a poll to name a new Mountain Dew flavor produced leading options like ``Hitler did nothing wrong'' \cite{noauthor_mountain_2012}, forcing the organizers to take down the naming process. For more, see \cite{rogers_boaty_2016}.} Looking ahead, collective input on the design and deployment of AI systems is an emerging special case for soliciting representative opinions, free from sockpuppet manipulation.\footnote{Some AI providers have commissioned research on how to best channel popular input in their tools, including via ``Alignment Assemblies'' to better align models with democratic will \cite{zaremba_democratic_2023, eloundou_democratic_2024, collective_assemblies_nodate, anthropic_collective_nodate}.}

\subsection{Mitigate bot attacks}
\label{ssec:mitigate_bot}

\begin{scenario}[ht]
    \centerline{\includegraphics[width=.9\linewidth]{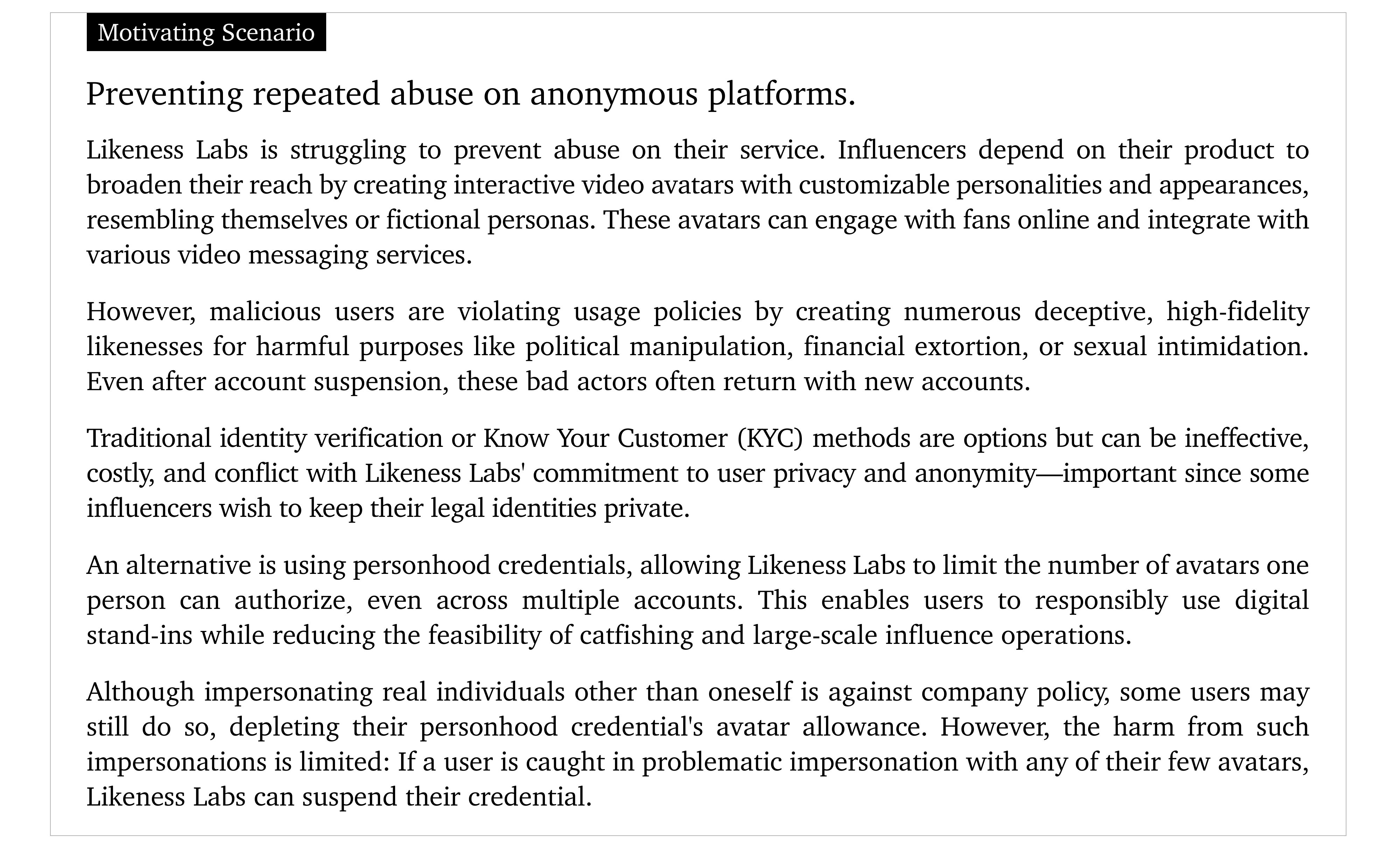}}
    \caption{Preventing repeated abuse on anonymous platforms.}
    \label{scenario:mitigate_bot}
\end{scenario}

Malicious actors use botnets---coordinated groups of entities meant to appear as distinct users---to carry out more effective attacks online. Sometimes, overwhelming a Web service with a number of distinct-appearing users is the point of the attack, as in distributed denial-of-service (DDoS) attacks \cite{silva_botnets_2013}. Other times, the purpose of relying upon many distinct-appearing users is to create many independent chances of success, like circumventing a service’s rules and creating new accounts for abuse even after having been caught. As discussed in \Cref{sec:introduction}, without new mitigations, malicious actors powered by AI systems will be more likely to succeed at these attacks.

Personhood credentials can help reduce the impact of bot attacks by enabling service providers to suspend users whose accounts have engaged in abuse. Today, adversaries’ tactics---like changing IP addresses and spoofing different browser configurations---often successfully evade the restrictions in place, at least for a period \cite{noauthor_game_2016}. Some existing software does guard against these attacks, but the developers of cybersecurity software have been locked in a perpetual cat-and-mouse game with attackers. And now, AI changes the dynamic, powering the attackers by reducing the manual work needed to get a new botnet running again after a previous one was caught. By using personhood credentials, service providers can restrict each individual to a limited number of accounts. This means that if someone breaks the rules and their account is suspended, they—identified through their personhood credential used at registration—cannot create new accounts without limit.\footnote{When establishing restrictions on user activity, websites must balance the level of abuse they are willing to tolerate against the potential loss of legitimate user engagement that such limits might cause \cite{mckenzie_optimal_2022}.}

Bot attacks have wide-ranging harms. Genuine users of peer-to-peer digital services are taxed when bots are not kept in check---they face increased friction and are targets of scams.\footnote{For instance, a website may have particular processes---phone number verification, rate limits---that are primarily meant to limit abuse by repeat bad actors but that inadvertently create difficulties for benign users.} Digital communities that aim to allow any individual to claim some asset (e.g., free trial memberships, credits for computational costs, products for beta testing), but only once, are often undermined by bot attacks. PHCs can serve as a valuable tool for checking who has already received the benefit, without revealing or tracking any further information.\footnote{A prosaic example of such fraud is free trials for digital services, which are often serially abused \cite{dixon_ditching_2020}.} This rate-limiting can also be applied to mitigate other forms of abuse that are exacerbated by rapid, AI-driven actions, such as ticket scalping, which disadvantages ordinary buyers \cite{ticketmaster_arms_2023, office_of_public_affairs_justice_2021}. Similarly, fraudulent AI-powered requests for benefits from governments or aid-giving organizations—even when successfully defended against—can drain resources and make it more difficult to serve the targets of the aid; screening applicants by PHC verification may help.\footnote{One risk is that screening via PHCs may reduce fraud but also make it more difficult for legitimate recipients to receive aid: Administrative overhead to apply for benefits tends to result in many people not receiving benefits that they are properly eligible for \cite{herd_introduction_2023, schweitzer_how_2022}.}

Looking ahead, we expect that many forms of AI-powered deceptive behavior that are especially harmful at scale can be \textit{indirectly} mitigated by personhood credentials. In the vignette in \Cref{scenario:mitigate_bot}, for example, we discuss one particular form of AI-powered deception that is harmful in isolated incidents but especially harmful at scale---impersonation.\footnote{By impersonation, we mean taking on the likeness of someone in particular, as opposed to taking on a sockpuppet persona---a plausible human that could exist but does not.} We dwell on this case because it nicely delineates how PHCs can and cannot directly help when it comes to AI-powered deception online: PHCs by default cannot validate a holder’s legal identity, and so a digital likeness service cannot directly check a PHC to confirm that the holder is taking on their own likeness rather than another person’s.\footnote{Malicious actors may have incentives to impersonate a real person to carry out social engineering attacks or to set up a profit-making scheme (e.g., if the target of impersonation is a celebrity).} However, such activities can be prohibited by digital services’ policies, and if a user is ever found to be in violation of the service’s policies, PHCs can make the policies more enforceable, by suspending the PHC-linked account and disallowing future signups for that PHC.\footnote{Beyond violating the policies of a digital service, using AI for impersonation might also sometimes involve breaking a law. Courts have sometimes ruled in favor of a ``right to publicity'' and ``right to privacy,'' which using a person’s likeness without their permission may violate, depending on the circumstances \cite{white1992, fraley2011}.} We expect that—anticipating these consequences—fewer attackers will attempt such unauthorized impersonations.

\subsection{Allow verified delegation to AI agents}
\label{ssec:verified_delegation}

\begin{scenario}[ht]
    \centerline{\includegraphics[width=.9\linewidth]{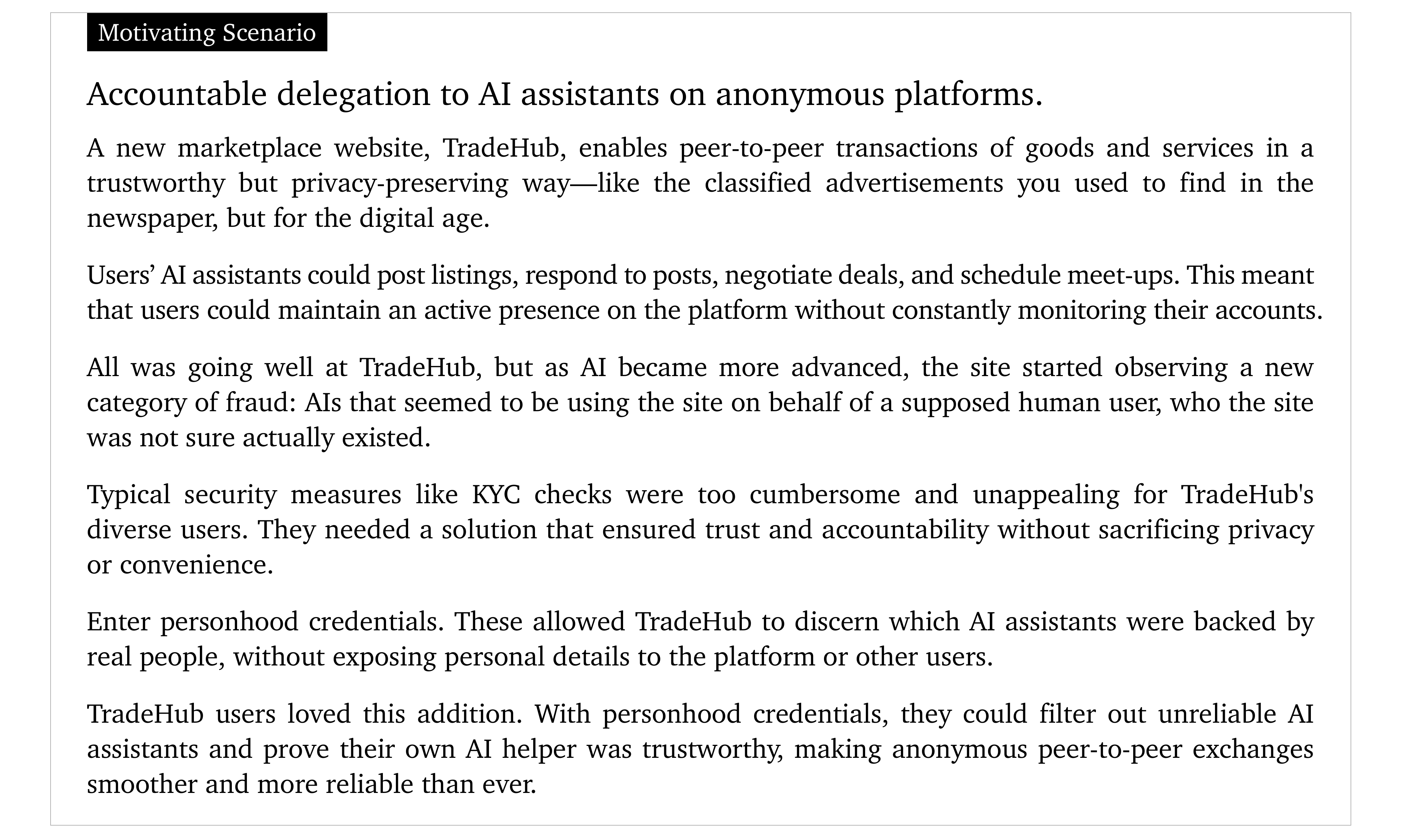}}
    \caption{Accountable delegation to AI assistants on anonymous platforms.}
    \label{scenario:verified_delegation}
\end{scenario}

AI systems are increasingly capable of acting autonomously \cite{kinniment_evaluating_2024, wang_survey_2024, fang_agents_2024, yao_react_2023}. While enabling many beneficial use cases,\footnote{For a description of some economically useful properties of AI agents, see \cite{shavit_practices_2023}. Already, some humans are deferring to AI-powered solutions for navigating dating apps on their behalf \cite{sengupta_russian_2024}.} this autonomy (``agenticness'') also facilitates a new form of deception: bad actors can deploy AI systems that, instead of pretending to be a person, accurately present as AI agents but pretend to act on behalf of a user who does not exist. This strategy exploits the current lack of norms around disclosure of agentic AIs, including a lack of norms around disclosing the identities of the people controlling them (often called their ``principals'').\footnote{In some jurisdictions there is already regulation governing bot disclosure \cite{kohne_californias_2019}. In 2023, legislation was introduced to codify bot-disclosure federally, though it has not become law \cite{noauthor_schatz_2023, brian_schatz_ai_2023}. Note that these bills outline requirements around disclosure of AI usage, i.e., requiring that people disclose when a bot was employed to take an action or when a piece of content was generated by an AI model. The outlined disclosure requirements stop there—they do not include disclosures related to the identity of the person or organization enlisting the AI.}

Personhood credentials could offer a way to verify that AI agents are acting as delegates of real people, signaling credible supervision without revealing the principal’s legal identity. This feature could be useful in a range of settings where users wish to rely upon AI assistants; \Cref{scenario:verified_delegation} outlines one such use case. Should a principal fail to address harms caused by their PHC-verified AI, they risk suspension from a service. Suspension implies that they lose their ability to verify delegates for some time period, reducing their capacity to perpetrate future harms.

Note how, in this case, PHCs create a form of accountability for AI agents without demanding sensitive information from principals.\footnote{A large and growing body of legal scholarship explores when and how humans should be accountable for the harms caused by AI agents \cite{etzioni_keeping_2016, chinen_co-evolution_2016, chinen_law_2019, Chesterman_2021, Chopra2024, lior_ai_2020, diamantis_vicarious_2023, kolt_governing_2024}. Many---though by no means all \cite{solum_legal_2020, forrest_ethics_2024}---of these proposals involve holding humans liable for some harms caused most directly by those agents. It is beyond the scope of this article to recommend any of these approaches over others or to explore the finer points of how such theories could work. We note, however, that these theories rely on the ability of someone harmed by an AI agent to sue the principal of that agent, which in turn depends on the principal being \textit{identifiable}---which a PHC alone does not achieve. Even without directly identifying the principal, however, PHCs can still shift the benefits of AI agent usage to be more positive. } This suspension mechanism may effectively signal which AI agents have trustworthy principals, even if verifying AI agents through principal-linked PHCs is voluntary. Agents that remain unverified might be perceived as having reasons for not undergoing verification.\footnote{This mechanism aligns with the ``unraveling result'' of voluntary disclosure models in economic theory \cite{Milgrom_good_1981, Milgrom_relying_1986}. In scenarios where parties hold verifiable private information---such as whether an AI agent operates under a trustworthy principal---even if revealing this information is optional, agents associated with trustworthy principals have a strong incentive to disclose it. Consequently, a lack of disclosure becomes informative: in equilibrium, agents that do not disclose are effectively signaling that they do not have trustworthy principals.} Moreover, some malicious activities involving autonomous AI agents may rely on hiding the fact that multiple agents are controlled by the same individual.\footnote{One example of how a person might enlist unverified agents to manipulate a market: A potential buyer on an auction site might create multiple personas to submit lowball offers, in hopes of baiting the seller into an artificially low assessment of their item’s value. In a similar example, a potential seller might enlist synthetic personas to create the appearance of higher demand for their item, inducing a legitimate buyer to pay a higher price and to act with more urgency. In each case, the ability to recognize these entities as linked to the same identity---even without knowing any further details of the identity---would help to head off these risks. For an estimation of the prevalence of these practices, known as ``shill-bidding,'' in online markets, see \cite{chen_how_2020}. For a discussion of market designs that are robust to these practices, see \cite{komo_shill_2024}.} PHCs can help address this issue by creating a framework that links multiple AI agents to a single principal without revealing the principal’s specific identity. By doing so, PHCs might make it more difficult for bad actors to conceal their network of AI agents, thereby reducing the potential for abuse that stems from undisclosed common principals.

Sometimes, a website---or a third-party user interacting with the agent---may need to verify the AI’s specific principal, not merely that it is backed by \textit{some} principal. Ultimately, a fuller framework for verifying AI agents \cite{chan_visibility_2024, chan_ids_2024} and their principals will likely be necessary.

\section{Prospective challenges for personhood credentials}
\label{sec:risks}

In this section, we discuss four areas in which PHCs' impacts must be carefully managed:
\begin{itemize}[noitemsep]
    \item \underline{Equitable access}: How might PHCs impact access to digital services?
    \item \underline{Free expression}: How might PHCs impact whether people feel safe and confident engaging across digital services?
    \item \underline{Checks on power}: How might PHCs change the power dynamics of digital services, in both substance and perception?
    \item \underline{Robustness to attack and error}: How might PHCs be vulnerable to mistakes and intentional subversion by different actors in the PHC ecosystem?
\end{itemize}

We highlight these areas as most salient among our discussions with security researchers, civil liberties groups, and builders of digital identity systems, and acknowledge that they cover only a small subset of all possible unintended consequences. This section is intended to be taken only as a preliminary outline. The risks discussed in this section can be managed through the pursuit of ideas discussed in \Cref{sec:next_steps}.

\subsection{Equitable access}
\label{ssec:reduced_accessibility}

Benefits from personhood credentials must be weighed against potential impacts on the accessibility of digital services. A need for frequent authentication of one's PHC could contribute to friction and frustration for users,\footnote{Although much of this paper focuses on the importance of preserving the Internet’s usability for \textit{people}, many automated software scripts are also important for a well-functioning Internet. Such scripts could be inadvertently affected by the widespread introduction of PHCs; websites may want to consider approaches that offer benign bots a way to proactively distinguish themselves. One example of how some websites communicate their preferences for bot interaction is the use of robots.txt files, which have existed in some form since the 1990s as an opt-in approach for contending with the impacts of bots navigating the World Wide Web \cite{koster_important_1994, cihonchilling}.} particularly those less comfortable with technology generally (like older adults, who are common targets of AI-powered deception \cite{the_white_house_fact_2023-1}).\footnote{It is well recognized in commercial contexts that the friction of authentication is typically a negative experience for users \cite{gavigan_measuring_2022, goodman_digital_2020}. Depending on how authenticating with a PHC works, ``Sign-in with PHC'' options may ease some of this friction.} If friction from PHCs is too high, some users may neglect to use them and thus lose out on the range of benefits they afford.

Relatedly, if service providers use personhood credentials to limit access to particular services, some groups of individuals without PHCs may be systematically excluded. It is thus important that PHC issuers design their systems with equitable access in mind, and that authorities guiding PHC ecosystem development are careful to ensure that all groups have some form of access to vital services.\footnote{One analogous precedent may be found in the United States’s Blueprint for an AI Bill of Rights, which states that certain vital services should provide a way for users to opt out of AI interaction; see ``Human Alternatives, Consideration, and Fallback'' in \cite{office_of_science_and_technology_policy_blueprint_nodate}.} Depending on the details of a PHC's implementation, some groups may have access before others:\footnote{For example, if PHCs are primarily backed by government IDs, there will likely be geographic disparity in access to PHCs initially. Around 850 million people worldwide do not currently have an official identification document \cite{world_people_2023}. Refugee populations, in particular, may struggle to obtain or maintain a PHC. This issue must be investigated, perhaps under the purview of existing programs studying refugee documentation, e.g., within the United Nations \cite{united_guidance_nodate}.} It is important that this not contribute to the economic exclusion of slow-adopting groups in the meantime.\footnote{PHCs will be less likely to contribute to economic exclusion if they are an opt-in mechanism for additional components of a service, rather than mandatory for the baseline service itself.} One positive possibility is that PHCs may help to unlock digital services for many who do not have access to forms of digital identity and are thus currently excluded from many basic digital services.\footnote{It is estimated that 3.3 billion people worldwide do not currently have access to an official digital identity for online transactions---PHC systems could offer an easier-to-access alternative \cite{world_people_2023}.} When assessing PHCs' relative impact on accessibility, it is also important to consider that as AI systems become increasingly indistinguishable from real people online, individuals and service providers may default to mistrusting digital strangers, which could likewise disproportionately affect some groups. 

To minimize their impacts on accessibility of digital services, PHCs must be robust to changes in a person’s circumstances. For instance, PHCs must be robust to changes in citizenship or residence. There must be fallback processes for PHCs that are not too taxing on the user, and which are extensively trialed to identify edge-cases. Further, PHC providers should recognize that a fear of exclusion due to life changes may be psychologically taxing, even if there are appropriate measures to regain access.\footnote{Processes for changing one’s name, gender, and other details can be very cumbersome. For instance, in the United States, a person must contact a host of government agencies when changing their name \cite{noauthor_how_2024}.}

\subsection{Free expression}
\label{ssec:reduced_expression}

It is important to consider how PHCs could reduce people’s willingness to speak freely and to dissent in digital spaces. Individuals may fear that their online speech will be linked to their offline identity, even though PHCs do not convey one's identity.\footnote{PHCs therefore respond to key criticisms of some platforms’ ``real name'' policies \cite{boyd_politics_2012, york_case_2011}.}  Both perceived and actual anonymity are important for expressing views without fear of retribution, particularly for marginalized groups or for those living under authoritarian regimes with extensive surveillance \cite{polyakova_exporting_2019, lai_examining_2022}.\footnote{Anonymity has been referenced as an essential part of public discourse. The United States Supreme Court stated in \textit{McIntyre v. Ohio Elections Commission} that anonymous speech ``is not a pernicious, fraudulent practice, but an honorable tradition of advocacy and of dissent'' \cite{noauthor_mcintyre_1995}. See also \cite{kosseff_united_2022}.}

PHCs also change the dynamics of credibility online, possibly in unpredictable ways that could affect individuals’ willingness to speak openly. Individuals who choose not to verify their personhood through PHCs might find their contributions discounted or even labeled as disinformation.\footnote{This dynamic has some similarity to the concept of the liar’s dividend \cite{schiff_liars_2022}. Though PHCs are implemented with the intent to mitigate online disinformation, they might have the side-effect of making non-PHC verified speech easier to write off as disinformation, even if it is in fact authentic.} On the other hand, statements that might previously have been dismissed due to doubts about their origin could earn credibility when accompanied by PHC verification. In some cases, hostile governments might be able to suppress authentic and credible dissent by limiting access to PHCs within their borders.\footnote{There are a number of ways that a hostile government might try to stop dissidents from getting PHCs. For instance, if a PHC system relies upon proprietary hardware, a hostile government might be able to restrict the hardware from its country, or otherwise control access to the hardware so that the government gains control over users’ PHCs \cite{buterin_what_2023}. Governments could also apply these prohibitions more narrowly to certain groups.} 

While PHCs preserve user privacy via unlinkable pseudonymity, they are not a remedy for pervasive surveillance practices like tracking and profiling used throughout the Internet today.\footnote{Such practices are widely documented; for one comprehensive treatment, see \cite{zuboff_age_2020}.} Although PHCs prevent linking the \textit{credential} across services, users should understand that their other online activities can still be tracked and potentially de-anonymized through existing methods \cite{eckersley_how_2010}. It is important to recognize that if PHCs are challenging or inconvenient for users to use, companies may be inclined to direct them toward alternatives that are easier to use but provide less privacy. Consequently, some users might opt for these more user-friendly, yet less private, solutions.\footnote{This is already seen with some examples of digital identity protocols and cryptocurrency wallets \cite{korir_empirical_2022}.} Emergent social dynamics may also threaten the anonymous nature of PHCs---for instance, if online abuse continues to be persistent even among individuals with PHCs, people may feel socially pressured to reveal auxiliary personal information to demonstrate that they are trustworthy.

There are many nascent legal and ethical discussions centered on the rise of highly capable AI and freedom of expression. It will be crucial to understand how PHCs fit into the legal and ethical frameworks that emerge. For example, one important legal question is when individuals need to disclose their use of AI \cite{kohne_californias_2019, noauthor_schatz_2023, brian_schatz_ai_2023}. Service providers might wish to allow AI delegation on their platform only with proper transparency to other users, or they might wish to disallow AI delegation on their service altogether.\footnote{For example, a dating application may wish to disallow users delegating any conversations to an AI agent if they have a ``verified person'' account. Nonetheless, the service provider may have a difficult time determining that the user has illicitly delegated to an AI agent, if this delegation occurs after signing up using one’s PHC. If services wish to curtail delegation---particularly when users have incentives to illicitly delegate---services will need methods for detecting signals of AI-powered behavior (e.g., if a user is on the service at all hours of the day) and for enforcing rules against it thereafter.} PHCs could provide one method for disclosing these new forms of expression, as well as a means of enforcing the related rules that do emerge.

\subsection{Checks on power}
\label{ssec:concentration_power}

One significant challenge for a PHC ecosystem is how it may concentrate power in a small number of institutions---especially PHC issuers, but also large service providers whose decisions around PHC use will have large repercussions for the ecosystem.\footnote{\Cref{apx:ecosystem} contains further discussion about how different design choices at the ecosystem level may produce power asymmetries between users and issuers, as well as between users and service providers.} Such challenges may be particularly acute in cases where the PHC issuer is not a democratically elected government with checks on its powers and accountability to its constituents, or where influential service providers are operating with little regulatory oversight. 

User information is one potential source of power for an issuer: Issuers should give strong assurances about what information they hold and for what reasons, favoring an ``as little as possible'' approach and confirming compliance via mechanisms like audits. Transparency over technical aspects of the system (such as open-sourcing certain software components\footnote{For example, Signal’s server \cite{noauthor_signalappsignal-server_2024} and protocol libraries \cite{noauthor_signalapplibsignal_2024} are open-source, and their technical specifications are publicly available (see, e.g., \cite{marlinspike_x3dh_2016}).}) may also reduce the risk that data are misappropriated\footnote{Even companies with large financial consequences at stake have suffered breaches of sensitive user data \cite{franceschi_23andme_2023}.} and the perception of such risks. Possible mitigations may include granting users a ``right to be forgotten'' \cite{wolford_everything_nodate, jones_ctrl_2016} by a PHC system—for instance, if they have lost trust in the issuer. 

The range of decisions under issuers’ and service providers’ purview may also create imbalances of power relative to users.  An issuer may be able to choose which use-cases and service providers to support;\footnote{In some identity systems, like India’s Aadhaar, private-sector service providers need to apply to the government to be authorized to identify people via their Aadhaar number \cite{UIDAI_requesting_nodate}.} similarly, service providers may be able to choose which issuers of PHCs to accept as legitimate. These choices might be made for reasons related to private incentives, rather than the well-being of users or an overall community. Furthermore, service providers may have opportunities for illegitimate exercise of their power through their increased ability to enforce controversial policies.\footnote{Just as services can use PHCs to stop bad actors from circumventing suspensions via new accounts, such enforcement could also impact benign users who have accidentally violated a service's rules. For services that use PHCs to enforce their rules, it is particularly important that users can discover and understand services' rules ahead of time.}

Proper democratic oversight, accountability, and transparency mechanisms must be in place to check the power of issuers and service providers, whether the issuers are governments or nongovernmental entities.\footnote{In some instances, governments have restricted data collection by issuers of nongovernmental proof-of-personhood systems like Worldcoin, claiming that data collection practices violated local laws \cite{njenga_global_2024} (some restrictions have been subsequently lifted \cite{worldcoinkenya}).} More broadly, the incentives of the issuer must be carefully considered—for instance, some existing PHC systems are associated with particular cryptocurrency tokens,\footnote{Historically, one motivation for proof-of-personhood systems—closely related to PHCs—has been to aim for equitable representation when voting in digital communities, like those associated with blockchain communities. Prominent examples of proof-of-personhood protocols with an associated currency include Idena, Worldcoin, Proof of Humanity, and BrightID \cite{decentralistcom_proof_2023}. A fuller evaluation of existing proof-of-personhood systems, and whether the credentials they issue meet the PHC requirements (credential limits and pseudonymous unlinkability), is beyond the scope of this paper.} which may introduce complex financial relationships between the issuer and credential holders.

\subsection{Robustness to attack and error}
\label{ssec:attack_error}

A PHC system, like any digital system, is vulnerable to attacks and exploits by multiple actors---most notably, subversion by the issuer itself, by service providers, and by users with malicious intent (for instance, those in a network of cybercriminals). Many cybersecurity best practices will apply to these systems---for instance, defenses against denial-of-service attacks that halt the issuance of new personhood credentials, and means to stop attackers from gaining access to sensitive records.\footnote{The potential harm of data extraction attacks can be reduced through strict adherence to the requirement, detailed in \Cref{ssapx:minimal_disclosure}, that PHC issuers store only the minimum necessary information.}

One important threat to consider is whether a PHC system is robust to subversion by its issuer. For instance, issuers may have asymmetric knowledge about vulnerabilities in their system, which allow them to issue fraudulent credentials for self-interested uses. One must also consider the possibility that a PHC issuer becomes less trustworthy over time---whether due to personnel changes or other factors. PHC systems may wish to commit to practices like third-party audit compliance to demonstrate their trustworthiness and ensure that users can discover if trust is no longer merited.

Another important threat to consider is how malicious service providers may use PHCs as cover to surreptitiously collect information about users. For instance, even though PHCs do not disclose any aspects of a user’s legal identity, a service provider might be able to trick users into uploading pictures of their government IDs to (purportedly) re-authenticate their PHC.\footnote{Standard authorization and re-authentication processes across services might reduce the ability for service providers to surreptitiously collect information through non-standard means.}

A final threat we consider is how users may try to subvert a PHC system to obtain multiple credentials, either at small scales or through coordinated attempts at cybercrime. These users could attempt to deceive the enrollment process, or could rely upon theft or illicit purchase of other people’s credentials.\footnote{One analog for these concerns is the market for resale of fraudulent accounts on social media sites \cite{thomas_trafficking_2013}.}

Beyond the attacks just described, a PHC system should also be robust to user error. For instance, PHC systems can offer methods to ensure that credentials can be recovered or revoked in case they are lost or stolen.\footnote{NIST offers recommendations on the management of cryptographic keys, which may be applicable to the design of recovery and revocation processes for PHCs \cite{joint_task_force_security_2020}. The easier that it is for a holder to recover or revoke a compromised PHC, the less incentive an attacker will have to try to compromise a credential, as they should not expect to control the compromised PHC for long. On the other hand, if it is too easy to revoke one’s PHC and be reissued a new one, this may reduce the issuer’s ability to enforce its credential limit requirement of one per person per issuer. We discuss this challenge further in \Cref{apx:implementation}.} A user should not be permanently locked out from vital digital services if they merely misplace their smartphone or other means of accessing their PHC; it is incumbent upon system designers to make PHCs error-tolerant for benign users.

These threat models and considerations are by no means comprehensive; we encourage research that classifies and assesses the many attacks that PHC systems may face, which will help inform the design of more robust systems. 

\section{Next steps for consideration}
\label{sec:next_steps}

Governments, technologists, and standards bodies---with frequent feedback from the public---can take steps to manage the risk of AI-powered scalable deception. Here, we outline two broad categories of ideas: \textbf{adapting} existing digital systems to prepare for the impacts of highly capable AI, and \textbf{prioritizing} the development of personhood credentials as one specific countermeasure.

\begin{table}[ht]
\centerline{\includegraphics[width=.9\linewidth]{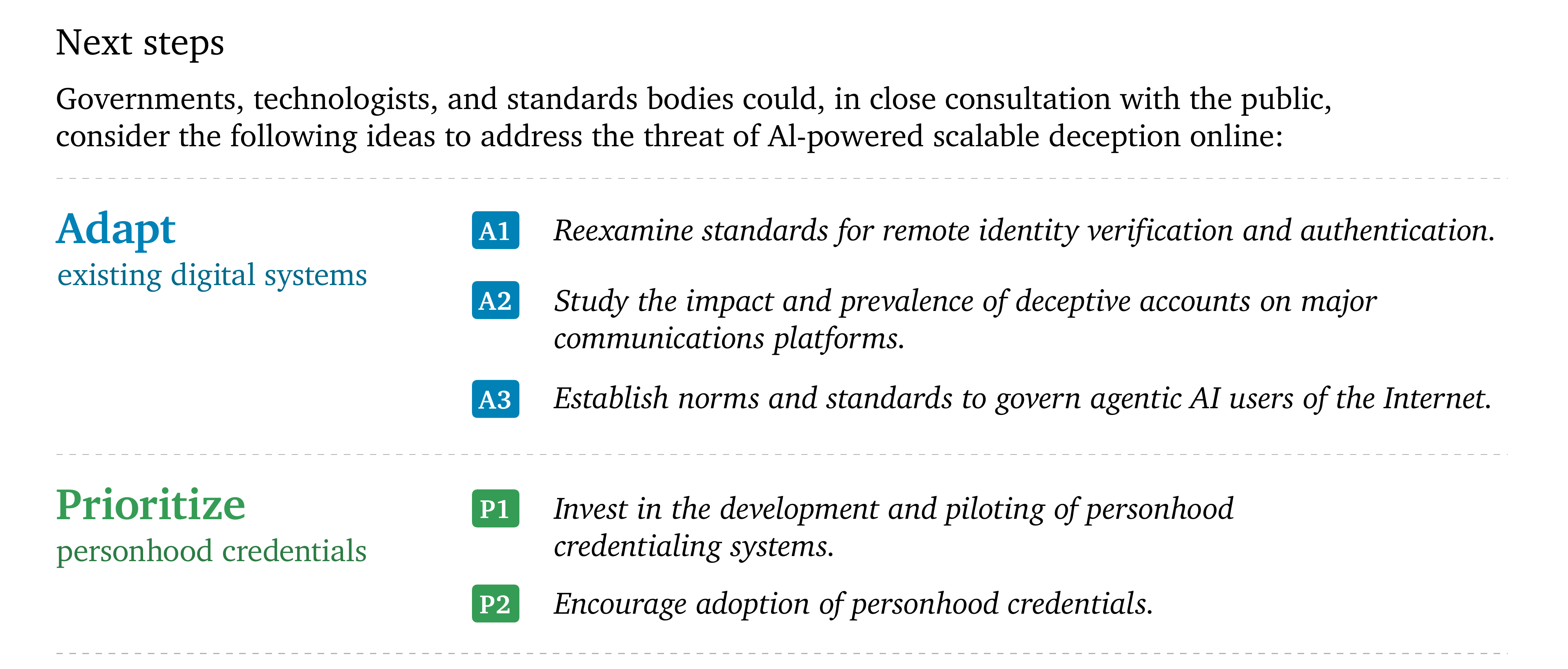}}
    \caption{Summary of next steps to consider.}
    \label{table:next_steps}
\end{table}

\subsection{Adapt existing digital systems}
\label{ssec:adapt_systems}

As we have discussed throughout the paper, many features of the Internet and digital infrastructure may struggle to contend with malicious actors whose deceptive schemes are powered by increasingly capable AI. It is important to assess the Internet’s readiness for these AI systems and to adapt accordingly.

\subsubsection*{A1. Reexamine standards for remote identity verification and authentication.}
\label{sssec:reexamine_identity}

The emergence of highly capable AI technologies may render certain security assumptions inaccurate. Such assumptions may be common to a range of standards that governments and industry actors rely on. Key standards that may require reevaluation include NIST’s Identity Assurance Levels (US),\footnote{NIST’s Digital Identity Guidelines are actively being revised \cite{joint_task_force_security_2020, noauth_nistsp_nodate}.} the European Union Agency for Cybersecurity’s (ENISA's) practices for remote ID proofing,\footnote{ENISA’s 2024 report includes a full section on deepfake-powered attack vectors, though it also suggests certain indicators of deepfakes that we do not expect to persist---for instance, certain characteristics about their bitrate or frames per second \cite{noauth_enisa_2024}.} and the international standard ISO/IEC 29115 regarding entity authentication \cite{ISOIEC29115}.

Anti-spoofing recommendations are one component of standards that may now be outdated: Previously, asking a subject to replicate a certain pose on a live video call was reasonable evidence that they were a real person rather than a prerecorded video of one.\footnote{For instance, NIST’s guidance on Identity Verification suggests: ``In order to confirm the video stream is live and not pre-recorded, the operator may direct the applicant to move their head in specific ways, raise or lower eyes, or ask the applicant questions requiring response during the live capture video'' \cite{nist_identity_2020}.} But malicious actors can increasingly use AI to mimic actions like these with lifelike avatars \cite{korshunov_deepfakes_2018, bray_testing_2023, xu_vasa_2024, homeland_increasing_2021, horvitz_horizon_2022}.

The relative strength of different forms of evidence may also warrant reevaluation based on AI’s improved abilities. For instance, authentication methods once considered effective in reducing fraud---such as voice-based authentication—are now increasingly vulnerable to AI-driven attacks. Advanced voice synthesis technologies can replicate an individual’s tone and speech patterns using minimal data, turning voice authentication into a potential avenue for account takeovers \cite{openai_navigating_2024}. In some domains (e.g., images and videos), the reliability of a piece of evidence might vary according to whether there is proof that the media was captured directly from a camera.\footnote{Photos taken with a C2PA-signing camera, like the smartphone application Truepic, can attest to being untampered \cite{truepic_authentic_nodate}.} More generally, reliability assessments of different forms of evidence may be improved by directly factoring in how AI might be used to subvert the evidence.\footnote{For instance, NIST considers biometric verification to be its highest strength of evidence, even if done via remote identity proofing \cite{nist_identity_2020}. To remain confident in the strength of this evidence, standards bodies should consider explicitly stress-testing any biometric algorithms for robustness to deepfake attacks: As NIST notes, ``The biometric False Match Rate (FMR)...does not account for spoofing attacks'' \cite{grassi_special_2023}.}

\subsubsection*{A2. Study the impact and prevalence of deceptive accounts on major communications platforms.}
\label{sssec:study_deceptive_accounts}

Currently, there are no standardized, official methodologies or tools to reliably measure the prevalence of AI-powered accounts across various digital platforms.\footnote{Without a standard approach, some have speculated that individual platforms have an incentive to permit some AI-powered accounts to drive up the total number of users and engagement, thus allowing the platforms to charge more money for advertising. For instance, ahead of Elon Musk’s acquisition of Twitter, he accused the site of engaging in these practices \cite{dar_we_2022}. The company that Musk commissioned to conduct a study of Twitter’s bot activity---Cyabra---has claimed to have benchmarks across social media companies, but to our knowledge has not made these data or methodologies available \cite{duffy_elon_2022}.} Developing such tools could enable platforms and governments to perform robust comparative analysis and to determine the effectiveness of different mitigation strategies. Encouraging independent auditing firms to assess the presence of AI-generated content and accounts might promote transparent and unbiased analyses across communications platforms. Moreover, restoring or enhancing researchers’ access to platform data can offer insights into the proliferation and trends of AI-driven accounts.\footnote{Some social media companies have phased out these forms of researcher access programs in recent years. In 2023, Twitter (now X) ended free access to APIs that had enabled research such as estimates of bot prevalence on the website \cite{calma_twitter_2023}. More generally, some social media companies have taken action to restrict research programs studying dynamics on the sites \cite{bond_nyu_2021}.}

Governments might explore several approaches to increase the availability of this information. For example, they could establish incentive programs for entities that share relevant information or that fund research grants focused on this topic. Clearly, the scope of such government activities would need to operate within existing legal protections of speech on private platforms,\footnote{In the United States, Section 230 of the Communications Decency Act is especially relevant when considering platforms’ responsibility for speech on their sites \cite{benson_section_2024}.} as well as user expectations of privacy.

\subsubsection*{A3. Establish norms and standards to govern agentic AI users on the Internet.}
\label{sssec:establish_norms}

We anticipate that, in the near future, AI agents will constitute a much larger portion of activity on the Internet than they do now \cite{cheung_mark_2024}. Appropriate trust in these agents can be facilitated by systems for confirming agents’ identities, properties, and claims (e.g., about which data and applications they have permission to access) \cite{chan_visibility_2024}.

As AI agents begin to interact with digital services in new ways, it is important to establish safeguards and guidelines around permissible interactions.\footnote{For suggestions of how to govern increasingly agentic AI systems, see \cite{shavit_practices_2023}.} Linking AI agents to personhood credentials could be a starting point, but a more comprehensive agent certification framework may be necessary for certain applications.\footnote{The development of HTTPS infrastructure may serve as an informative case study. The HTTPS certificate ecosystem uses a series of trusted institutional checks and public-key cryptography to allow users to trust that they have been routed to the correct website \cite{durumeric_analysis_2013}.} Many organizations are already working on best practices for issuing, authenticating, and presenting digital credentials---the EU Digital Identity Wallet Consortium \cite{eudi_arf_2024} is a prominent example---and these initiatives can more directly prepare for highly capable AI.

\subsection{Prioritize personhood credentials}
\label{ssec:prioritize_personhood}

We encourage governments, technologists, and standards bodies to prioritize the development, piloting, and adoption of personhood credentials (PHCs) to reduce AI-powered scalable deception.

\subsubsection*{P1. Invest in the development and piloting of personhood credentialing systems.}
\label{sssec:invest_credentialing}

To be useful, PHC systems must attain a meaningful scale.\footnote{There are some public estimates of the scale of proof-of-personhood systems, which are closely related to PHC systems. The largest of these, Worldcoin, announced in July 2024 that it had reached 6 million signups \cite{worldcoin_just_2024}. Idena, another proof-of-personhood system, appears to have had on the order of several thousand members as of April 2022 \cite{idena_2021, ohlhaver_compressed_2024}. The proof-of-personhood system Proof of Humanity had roughly 17,000 members as of 2022 \cite{merk_ethnographic_2023}.} One way to build trust and spur adoption is to build PHC systems incrementally atop existing government credentials, such as driver’s licenses and passports.\footnote{Such work might fit naturally into an ongoing initiative led by NIST in the US related to digital driver’s licenses and the management of digital identity on mobile devices generally; see \cite{mehta_accelerate_2023} for write-up and \cite{noauthor_digital_nodate-1} for project status. Certain types of digital driver’s licenses have different privacy properties, which may afford different functionality when used as a basis for a PHC. For criticism of the mDL standard (ISO/IEC 18013-5) for digital IDs, see \cite{stanley_tsa_2023}. A digital ID's validity period is one factor that might affect the potential ability to triangulate users' activity within a certain time window, and therefore affect the ability to satisfy the PHC requirement of pseudonymous unlinkability \cite{baum_cryptographer_nodate}.} We expect that public-private partnerships will be important for moving nimbly: A number of technologists have already developed proofs-of-concept for converting government identification documents into privacy-preserving credentials\footnote{For two examples of protocols that convert an existing identity document to a privacy-preserving credential, see \cite{noauthor_zk-passportproof--passport_2024,AnonAadhaar}. More broadly, the US government has funded infrastructure solutions that demonstrate support for privacy-preserving digital credentials \cite{st_public_affairs_news_2023}.} and may be amenable to expanding these pilots. Notably, some regional governments, such as the government of British Columbia, are already issuing similar credentials \cite{british_person_2024, bc_verfiable_2021}. If many governments or multinational organizations pursue PHCs, a supranational entity could help to coordinate PHC systems internationally.\footnote{Such an entity could focus on cross-border interoperability and standards, akin to how the UN’s International Civil Aviation Organization manages standards for biometric passports \cite{noauthor_icao_nodate}.}

Accelerating research and development of PHC systems---particularly their privacy, inclusivity, and fraud-resistance---is essential. The defenses of PHC systems can be assessed through red-teaming exercises that identify vulnerabilities and point toward countermeasures \cite{zenko_red_2015}. Inclusivity research might benefit from exploring the adoption barriers faced by various demographic groups.\footnote{For example, policy initiatives related to broadband access may serve as an instructive precedent---in that case, bridging the digital divide required focusing special attention on rural communities, people with disabilities, and underserved groups \cite{gao_closing_2023, kelly_broadband_2020}.}

There are many open questions, particularly legal ones, surrounding personhood credentials:
\begin{itemize}[noitemsep,topsep=0pt]
    \item How should personhood credentials relate to existing identity theft and protection laws?
    
    \item In scenarios where service providers currently cannot require government ID, should similar restrictions apply to government-issued personhood credentials?\footnote{As an example of restrictions on existing government ID programs, India’s Supreme Court ruled that its banks may not demand Aadhaar details as a mandatory process for opening a bank account \cite{manveena_aadhaar_2018}.}
    
    \item Are there specific use-cases that warrant granting individuals the right to maintain multiple verified pseudonymous accounts, as opposed to allowing service providers to mandate a single verified account per person?

    \item How can users confirm that privacy commitments from an issuer and service provider are being upheld, without relevant cryptographic expertise?
\end{itemize}

Research that explores these issues will help stakeholders evaluate implementation trade-offs. Moreover, people can then engage in substantive dialogue that enhances PHCs through broad-based deliberation.

\subsubsection*{P2. Encourage adoption of personhood credentials.}
\label{sssec:encourage_adoption}

Encouraging service providers to accept personhood credentials (PHCs) as alternatives to traditional ID verification---especially in scenarios where legal identity is not strictly necessary---could substantially boost PHC adoption and enhance privacy.\footnote{For an illustration of the risks of sharing government IDs for verification purposes, see \cite{cox_id_2024}.} Governments might also define situations where individuals have a right to interact with only other real people, highlighting PHCs as one tool for achieving this.

To foster PHC demand, governments can promote their use in digital civic activities, like public comment processes. These uses not only normalize PHCs but also showcase their practical benefits, fostering broader public acceptance and trust. Additionally, governments can consider offering subsidies or incentives targeting user enrollment and inclusive outreach efforts.

To facilitate PHC adoption, it is valuable to improve general digital accessibility and literacy. Pre-installing digital credential management applications on smartphones and highlighting them as key features can aid users.\footnote{The ability for iOS and Android devices to store digital identity cards in their Wallet applications is one possible start for this \cite{support_add_nodate, noauthor_store_nodate}.} Encouraging broader use of tools like password managers can prepare individuals to secure their personhood credentials effectively.\footnote{Greater understanding of concepts like encryption and zero-knowledge proofs may also help people understand the ways their personhood credentials are secured—easing some anxiety about whether PHCs’ use is truly private.} As users become more comfortable with these technologies, they are more likely to enroll in PHC systems and to use PHCs successfully.

\newpage

\section*{Acknowledgments}\label{sec:acknowledgments}
\phantomsection
\addcontentsline{toc}{section}{Acknowledgments}

This research project and its resultant paper---focused on the challenges that AI could pose for trustworthy digital interaction, and on prospective solutions to these---are the efforts of a large multidisciplinary collaboration, across researchers at many different organizations, with many different points of view and equities. For instance, coauthors hold a range of views on the relative merits of different “roots of trust” for a PHC---whether to build on government IDs, Webs of Trust, and/or biometrics. These disagreements center on complex empirical and normative questions around the robustness, inclusivity, and privacy risks of different methods. Likewise, coauthors hold a range of views regarding whether an ideal PHC system should offer enrollment to people worldwide or focus more narrowly on a certain geography. One objective of this paper is to catalyze public discussion and debate about these topics. This paper should not be read as implying the full point of view of any particular coauthor or of the organization to which they are affiliated. Further, this paper should not be read as advocating for or against any particular PHC system or digital credential.

We are grateful to a number of groups for feedback and input that helped to shape the ideas and presentation of this paper. We would like to thank Sandro Herbig of Tools for Humanity\footnote{Tools for Humanity is the technology company that led the initial development of Worldcoin / World ID---an instance of a proof-of-personhood system---and works to support the World ID project’s technology protocols. One of Tools for Humanity’s co-founders, Sam Altman, is also co-founder and CEO of OpenAI. This paper is not an endorsement of World ID or any specific personhood credential or proof-of-personhood system.} for his contributions to these ideas, including detailed explanations of proof-of-personhood systems’ functionality and common points of feedback and confusion related to these systems. We would like to thank Eden Beck and Erol Can Akbaba for copyediting and typesetting, respectively, and for generally sharpening the presentation of the paper’s ideas.

We would like to thank the AI and Media Integrity Steering Committee of the Partnership on AI and the Berkman Klein Center for Internet \& Society at Harvard University for hosting us for feedback sessions related to these topics. We would also like to thank a number of individuals for feedback and conversations: Scott Aaronson, Mada Aflak, Dan Alessandro, shirin anlen, Rahul Arora, James Aung, Boaz Barak, Katie Benjamin, Jake Brill, Miles Brundage, Jon Callas, Rosie Campbell, Alan Chan, Melissa Chase, Peter Cihon, Paul Crowley, Karen Easterbrook, Tyna Eloundou, Emiliano Falcon-Morano, Jason Fedor, Claudia Fischer, Tim Fist, Lydia Gorham, Mark Gray, Margaret Hu, John Jordan, Gabe Kaptchuk, David Karger, Pedram Keyani, David Kim, Grace Kwak Danciu, Kim Laine, Michael Lampe, Jaron Lanier, Teddy Lee, Traci Lee, Brenda Leong, Leon Maksin, Timothy Marple, Stephen McAleer, Reed McGinley-Stempel, Pamela Mishkin, Ben Newhouse, Cullen O’Keefe, Christian Paquin, Joel Parish, Tobias Peyerl, Brendan Quinn, Ankur Rastogi, Francis Real, Elizabeth M. Renieris, David Robinson, Ben Rossen, Girish Sastry, Bruce Schneier, Eric Scouten, Gaurav Sett, Yonadav Shavit, Dane Sherburn, Adam Shostack, Allison Stanger, Jay Stanley, Esther Tetruashvily, Değer Turan, Raquel V{á}zquez Llorente, Chelsea Voss, Becky Waite, E. Glen Weyl, Rebecca Williams, Wenda Zhou, and Jonathan Zittrain.

\clearpage
\phantomsection
\addcontentsline{toc}{section}{References} 
\bibliographystyle{abbrvnat_mod2}
\bibliography{ProofofPersonhoodCollabReferences.bib}

\newpage
\appendix
\addappheadtotoc
\addtocontents{toc}{\protect\setcounter{tocdepth}{1}} 

\appsection{What do we mean by ``trustworthy'' digital interaction?}\label{apx:conceptual}

Extensive academic literature explores the concepts of trust and trustworthiness, in both online and offline settings---too many to cite here. In this paper, when we refer to ``trustworthy'' digital interactions, we mean interactions that deliver on parties’ reasonable expectations:\footnote{We do not aim to be exhaustive with this definition.} for instance, an online ecosystem in which service providers are confident that their services are being used as intended, and in which users can participate in digital services as intended, free from fears of abuse, attack, and other harms. Deceptive activity breaches trust---it is untrustworthy. 

To set and fulfill parties’ reasonable expectations, various contextual details about an interaction may need to be provided or interpreted by a party to the interaction: Who is on the other side of this interaction? What might their interests be in this matter \cite{jain_contextual_2024}? Untrustworthy or deceptive content like disinformation often depends upon giving one party a misleading impression about the context of that content.\footnote{Recently, a widely shared photo on X purportedly from the center of a conflict zone was in fact pulled from a video game \cite{novak_seven_2023}.} Scams, as another example, give misleading context to trick users into thinking they are engaging with a legitimate service.

Providing people with (and fulfilling) reasonable expectations of an interaction is not the only important value; it is critical to also safeguard privacy, which may be in tension with a desire from other parties to know more about the participant.\footnote{It is possible to imagine policies that lean too far in favor of one value or another: From a privacy perspective, it would be significant overreach to require strong forms of identification for any and all activity across the Internet. Likewise, a website might struggle to uphold trustworthy interactions if it is unwilling to collect or convey any information about its users to one another---though this might also be appropriate, depending on the website’s purpose.} One approach for balancing these conflicts is for contextual information to be calibrated to only what is necessary to carry out the interaction.\footnote{This view aligns with Helen Nissenbaum’s notion of privacy as \textit{contextual integrity}, outlined in \cite{nissenbaum_privacy_2009}. An information flow is private if it respects the contextual norms governing the interaction. More generally, this is related to the principle of ``data minimization'' \cite{edps_data_nodate}.}

As deceptive actors increasingly look to AI to execute their schemes, how do we ensure that users and service providers can verify details that properly contextualize their interactions, while restricting information disclosure to what is necessary?

Few existing approaches allow for a robust yet minimal verification of a type of information that will become increasingly important as AI capabilities continue to improve and AI tools expand their reach: proof that there is a real person behind some digital activity. From this perspective on what constitutes ``trustworthy'' digital interaction, personhood credentials are a natural tool to pursue. We view this pursuit as part of a broader push toward anonymous credentials \cite{chaum_security_1985} that prove minimal claims tailored to context (e.g., a digital credential that proves ``I am over 18 years old'' but nothing more).

\pagebreak
\appsection{Relating personhood credentials to CAPTCHAs and synthetic content transparency tools}\label{apx:current_approaches}

In this section, we further explore two common alternative approaches to the problem of scalable AI-powered deception online. PHCs can complement these approaches.

\subsection{\textit{CAPTCHAs and other behavioral filters}}
\label{sapx:captchas}

The most common behavioral filter for AI systems is CAPTCHA (Completely Automated Public Turing test to tell Computers and Humans Apart), developed in the early 2000s to differentiate humans from bots by presenting challenges that were too difficult for the bots of that era \cite{von_ahn_captcha_2003}.

There are various types of CAPTCHAs---such as recognizing obscured words or solving rotation-based puzzles---and there are also various methods to undermine them \cite{kumar_systematic_2021}.

Even before AI could directly solve CAPTCHAs, malicious actors found ways to use AI systems to bypass these barriers. For example, testers discovered that GPT-4---a text-only AI model lacking ``vision'' capabilities---pretended to have a disability to persuade a skeptical human to complete a CAPTCHA on its behalf \cite{openai_gpt-4_2024, noauthor_update_2023}. More broadly, it is easy for automated scripts to enlist human workers to solve these tests for fractions of a cent per CAPTCHA.\footnote{A bot calls on these human workers when it detects it has been presented with a CAPTCHA challenge, through services like 2captcha \cite{datadome_captchafarm_nodate}. For detailed reports, see \cite{woods_human_2021} and \cite{noauthor_cybercriminals_2017}.}

Now, AI systems have developed capabilities that further challenge the effectiveness of CAPTCHAs in filtering out bots online. Multimodal AI, which combines vision with cognitive abilities to perform various tasks, in particular, poses a significant challenge.\footnote{For a review of OpenAI’s GPT-4V and its current abilities to pass certain forms of CAPTCHA, see \cite{noauthor_gpt-4vision_2023}.}

As AI systems get closer to exhibiting a range of human-like abilities, it will get increasingly difficult to design CAPTCHAs that all people can easily solve, but AI systems cannot.

It is worth noting that for many service providers that use CAPTCHAs, the aim is not to make bot attacks impossible but to render them economically unviable. CAPTCHA deployers might consider their measures successful if they make abuse more challenging for bots, even if some bots occasionally succeed. Nevertheless, as AI becomes less expensive, the economic barriers imposed by CAPTCHAs diminish, while humans might face increased difficulty in solving more challenging tests.

Some digital services have moved beyond CAPTCHAs to employ a broader range of behavioral filters aimed at reducing bot-based deception. For instance, anomaly detection methods attempt to identify account activity patterns that appear suspiciously coordinated or unusual compared to that of the general user base.\footnote{For one example of anomalous behavior---unusual signup patterns---from fraudulent users, see \cite{yuan_detecting_2019}. One way that such accounts get discovered is through common patterns in their activity, which may be the result of idiosyncratic details in a malicious actor's automated script. AI agents might be able to vary their practices in ways that are harder to correlate, as opposed to following a relatively static script.} JavaScript-based browser challenges assess factors like the manner in which users access a service.\footnote{For an overview of such challenges and commercial providers of these, see \cite{aminazad_webrunner_2020}. One way that bots already attempt to appear more person-like is through taking actions like moving the mouse and page-scrolling, as well as through acting on a time delay. We expect these actions will become more natural-seeming as AI systems learn to better imitate human website behavior, through methods such as behavior cloning \cite{torabi_behavioral_2018} or other advanced techniques \cite{ho_generative_2016}, which can allow for learning complex behavior even when the behavior’s objective cannot be well-specified in advance. For an example of imitating relatively complex human behavior in digital environments, see \cite{baker_learning_2022}.}

These advanced methods fundamentally rely on discernible differences between how people and AI systems interact with a particular website. As AI systems become harder to distinguish from people, websites may face difficult trade-offs. To significantly reduce AI-powered abuse, they might need to limit a substantial portion of people’s activity---which cannot be reliably differentiated from AI-based traffic---in the process. Personhood credentials could offer an alternative that avoids this trade-off by offering a new way to identify automated activity that does not depend on outdated assumptions about AI systems’ sophistication.

\subsection{\textit{Synthetic content transparency tools}}
\label{sapx:synthetic_transparency}

Synthetic content transparency tools intend to help distinguish AI-generated (``synthetic'') content from content that people created without using AI. The Partnership on AI’s Responsible Practices for Synthetic Media project has outlined many such tools \cite{pai_staff_building_2023}, including:
\begin{itemize}[noitemsep, topsep=0pt]
    \item \underline{Watermarking}: proactively modifying AI-generated content (visibly or invisibly) to contain interpretable signals of how content was generated or edited.\footnote{An example of a visible watermark is the color palette applied to the bottom-right of images generated by DALL-E 2 \cite{noauthor_dalle_nodate}. For a discussion of invisible watermarking of AI-generated images, see \cite{noauthor_stable_2023}. For further technical details on watermarking in the domain of text, see \cite{Kirchenbauer_2023}.}

    \item \underline{Metadata (Provenance)}:\footnote{The Partnership on AI uses the term \textit{metadata} to refer to this mitigation, rather than the term \textit{provenance}, which sometimes is taken to mean a wider umbrella of mitigations for telling ``where content is from'' \cite{pai_staff_building_2023}. \textit{Metadata}, in contrast, refers specifically to ``labeling in metadata where the content is from.''}including information about content’s origin or its edit history, potentially with cryptographic signatures for tamper-resistance (as in C2PA—the most prevalent example of this method\footnote{The most prominent metadata provenance group is the Coalition for Content Provenance and Authenticity (C2PA), which aims to ``[address] the prevalence of misleading information online through the development of technical standards for certifying the source and history (or provenance) of media content'' \cite{noauthor_c2pa_nodate}. C2PA is supported by a number of technology and media companies, including Adobe, the BBC, Google, Microsoft, and OpenAI, and several technology companies have begun to use its standard for tracking AI-generated images \cite{clegg_labeling_2024, david_openai_2024}. Similarly, makers of physical cameras, like Sony, and of software for smartphone cameras, like Truepic, are developing tools that enable C2PA (or C2PA-like) tracking so that a photo taken with their equipment would automatically be associated with a C2PA manifest \cite{dulai_sony_2023, truepic_truepics_2022}.}).

    \item \underline{Fingerprinting}: logging AI-generated content into a database so that future content can be checked against this, generally through hashing.\footnote{This approach has some similarities to hashing illegal sexual material---such as CSAM---to detect it and thus reduce its spread \cite{noauthor_how_2023-1}.}

    \item \underline{Classifier-based detection}: using AI models to assess (predict) the likelihood that content was generated by AI.\footnote{For an overview of many methods described as ``classifier-based detection,'' see Texas Tech’s list of AI detection software \cite{quinn_artificial_2023}. For an example classifier that does not require training a new model for detection, see \cite{mitchell_detectgpt_2023}.}
\end{itemize}

There are two central reasons why personhood credentials are an important complementary tool for addressing scalable AI-powered deception online.

First, while there are promising developments in synthetic content tools’ adoption and efficacy, these tools are inherently limited in scope: Given that many AI models are and will likely continue to be open-weights, we expect it will not be possible to enforce the use of these tools in all models.\footnote{In some open-weights implementations of models, the watermarking function can merely be removed from the model’s code before running \cite{noauthor_stable_2023}.} In addition, a perennial problem with detection tools is that outputs from a model may be modified to evade detection.\footnote{Even if an invisible watermark is applied to AI-generated content, adversaries may have reliable enough methods for perturbing the content so as to remove this watermark \cite{zhang_watermarks_2023}. For further discussion of methods for defeating watermarks, see \cite{kelly_watermarks_2023}.} In the case of metadata-based provenance, one simple evasion is to detach a piece of content from its metadata.\footnote{Some websites strip metadata from images by default to protect their users’ privacy \cite{laurent_study_2013}. Groups like C2PA are well aware of the threat that adversaries may remove metadata from a piece of content \cite{noauthor_c2pa_nodate}. Including signed metadata can be a useful intervention despite users’ ability to strip this information: For instance, in due time, users might learn to be skeptical of images that are not paired with metadata attesting to their provenance from a physical camera. For metadata with an intention of disclosing that a piece of content was AI-generated, however, the ability to strip this metadata is a larger impediment.} Moreover, the ultimate value of these tools depends on designs that aid end users’ interpretations of the information, which are still open areas of research \cite{feng_examining_2023}. Despite these drawbacks, synthetic content transparency tools are valuable and worth pursuing; they just are not a sufficient solution on their own and are best pursued in concert with other mitigations.

Second, the challenge of AI-powered deception online is significantly broader than just synthetic content; PHCs focus on a complementary part of the challenge. In common domains of interest, like social media, AI-generated content is not strictly necessary for malicious actors to still carry out AI-powered deception. For instance, AI-powered accounts might choose to amplify posts from real humans that happen to support an agenda they wish to promote. Likewise, AI-powered accounts might be used to increase the clout of particular other accounts, such as by adding to their follower count or by simply liking their posts. In both cases, there is no AI-generated content for synthetic content transparency tools to detect. Understanding whether there are real people behind these accounts, on the other hand, is well-suited to mitigating AI-powered deception even in these cases. Furthermore, as AI systems become increasingly agentic---less like content-generating machines, more like full-fledged Internet users---we expect that AI systems will increasingly interact with websites in ways that are not amenable to content-based interventions, whereas PHCs will continue to be a valuable signal.\footnote{In the case of botnet attacks, for instance, malicious actors enlist AI to carry out repeated abuse of a website’s services---but if the bots are primarily interacting with the website through clicks and loading particular webpages, there is no content for the website to check for evidence of being AI-generated.} (We offer further examples below of cases in which AI-powered deception necessitates going further than synthetic content transparency tools.)

Because PHCs focus on a separate portion of the challenge from synthetic content transparency tools, the methods may be strongest as complements to one another. For example, content provenance manifests could incorporate personhood credentials as one type of supported identity attestation.\footnote{Building on the C2PA standards, the Creator Assertions Working Group (CAWG) is a separate organization that has built protocols that allow creators to make additional assertions about their content---most relevant to PHCs being its identity assertions \cite{noauthor_creator_nodate}. Through a CAWG identity assertion, a content creator can digitally sign a C2PA manifest so that their identity is attached to the content as it moves through digital space. Including evidence of a personhood credential in such an assertion could provide strong evidence of a human’s involvement in the content’s creation. CAWG has begun work on supporting Verifiable Credentials---a potential backbone of PHCs---as one form of identity assertion \cite{noauthor_creator_nodate}.} For creators looking to verify their involvement with a particular piece of content, this would allow them to do so without requiring that they attach their legal identity. More broadly, synthetic transparency tools and PHCs are complementary in the general cause of encouraging people to be more discerning about who and what they see online. As Internet users become more comfortable interpreting information from one set of tools and methods, it is possible that this level of literacy will extend to others.

To emphasize the ways in which synthetic content transparency is insufficient for solving some forms of AI-powered abuse, we offer a few illustrative cases in which we expect a different approach is needed. 

In some cases, AI-powered abuse can occur without relying on AI-generated content. This could occur when AI leverages human-generated content (\textbf{Case 1} and \textbf{Case 2}), or when AI interacts directly with websites (\textbf{Case 3}). In other cases, AI-generated content is central to the abuse, but transparency-centric approaches do not fully resolve the harms (\textbf{Case 4}). In yet other cases, AI-generated content is beneficial, and synthetic content transparency tools may be misinterpreted and make these uses less possible, if not complemented by other solutions \textbf{(Case 5)}.

\subsubsection*{Case 1: A bad actor uses AI to spread human-generated content for harassment.}

Alice wants to use AI to embarrass Bob: She uses AI to periodically create new accounts on social media, which post a real photograph of Bob doing something that she knows would embarrass him. Because the photo does not originate from AI, synthetic content transparency tools do not flag it, despite AI being used to amplify its reach.

\subsubsection*{Case 2: A bad actor uses AI to curate human-generated content to push a political agenda.}

  Alice wants to use AI to push a particular political agenda on social media: She uses AI to automatically analyze a stream of real people’s posts, with a network of accounts that repost the material AI determines best supports her agenda. Because the material in the posts does not originate from AI, synthetic content transparency tools do not flag it, despite AI being used to automate all reposting of content by Alice's accounts.

\subsubsection*{Case 3: A bad actor deploys AI to take actions on a website and search for exploits.}

Alice wants to find and exploit vulnerabilities on a website to demand a ransom payment: She uses a wide range of AI agents to analyze the site’s source code and to test out different sequences of unusual inputs. Because the agents interact primarily with the website through mouse-clicks and simple text-field inputs, synthetic content transparency tools are largely unhelpful in this case.\footnote{Synthetic content transparency tools will not address mouse-clicks and other forms of browser navigation, which are action-based rather than content-based. A website could attempt to look for watermarks in content submitted to its text fields, but watermarking tends to be unreliable for short pieces of text.} Moreover, the site receives only a very small portion of content that the AI has generated; it cannot analyze the AI’s private chain-of-thought in planning and executing this abuse.

 \subsubsection*{Case 4: A bad actor uses AI to generate abusive content, but identifying the content as AI-generated does not resolve the abuse.}

Alice wants to use AI to bully Bob by depicting him in images she knows he will find upsetting: She uses AI to create these images and then posts them to message boards she knows Bob frequents. Because the message board does not have a blanket ban on AI-generated images---and because disclosure of the images being AI-generated does not reduce the targeted harassment that Bob feels---synthetic content transparency tools do not stop this abuse, despite AI-generated content being intrinsic to it.

\subsubsection*{Case 5: AI empowers a good actor to engage with others digitally, but the good actor's AI-generated content causes them to be mistaken for a bad actor.}

   Bob lost the ability to speak with much fluency after a medical incident, but he was able to recover his ability to speak by using an AI tool. Now, Bob can once again speak with others over the phone---even with his own voice. Unfortunately, many companies have started to automatically flag and filter out callers they suspect are voiced by AI, given a surge of AI-powered phone scammers. Because Bob’s voice is in fact AI-generated, he is frequently flagged in these settings and has a harder time when he needs to call in for customer support. Without some alternative method, Bob struggles to prove that he is a real person---using AI to help him express himself---rather than a bad actor using AI to deceive.

\appsection{Implementation choices for personhood credentials}
\label{apx:implementation}

There are a number of ways that a PHC system can achieve the requirements we outline. As a reminder, those requirements are:
\begin{enumerate}[noitemsep, topsep=0pt]
    \item \textbf{Credential limits (1 credential per person per issuer):}
    \begin{enumerate}[a.]
        \item \underline{Issuers check one-per-person requirement at enrollment.}

        \item \underline{Expiry or regular re-authentication.}
    \end{enumerate}

    \item \textbf{Unlinkable pseudonymity (privacy):}
    \begin{enumerate}[a.]
        \item \underline{Minimal identifying information stored during enrollment.}

        \item \underline{Minimal disclosure during usage.}

        \item \underline{Unlinkability by default.}
    \end{enumerate}
\end{enumerate}

We next describe possible methods. We do not aim to give a comprehensive summary of all possible implementations here, nor do we intend to endorse any of the implementations that are discussed here. The goal of this section is simply to provide a brief overview of the range of methods that could be employed to achieve the requirements.

\subsection{\textit{Methods to achieve a one-per-person credential limit}}
\label{sapx:methods_limit}

\subsubsection*{\underline{Issuers check one-per-person requirement at enrollment}}
\label{ssapx:check_enrollment}

There are three main methods through which an issuer can ensure that a particular user has not received a credential from them before (and thus enforce a credential limit): existing identity documents,\footnote{Such documents are sometimes referred to as ``breeder documents'' in the security literature.} biometric information, and Webs of Trust (social graphs). These methods can also be combined to, in some cases, achieve higher assurance. 

Importantly, an issuer can choose a verification method that originates from a different source as long as the issuer trusts its reliability. For example, a nongovernmental issuer could issue PHCs built atop government IDs, rather than devising its own method for giving only one credential per person. Likewise, a government issuer could choose to issue PHCs atop a method other than an existing government ID. 

One option for limiting the number of credentials per person from an issuer is to rely upon existing forms of ID, so long as those systems have finite limits per person. For instance, relying upon birth certificates or tax IDs can achieve finiteness (a credential limit) if these forms of ID are also finite, though they may also have edge-cases; the issuer may need a further mechanism for deduplication in those cases. Moreover, just as some forms of ID are harder to acquire (e.g., passports) than others (e.g., library cards), a PHC provides stronger claims of being a real distinct person when it is backed by an ID that is difficult to procure in large quantities.

How can an issuer authenticate the existing form of ID, so that they do not accidentally issue personhood credentials based on a spoofed document? Some forms of existing ID are already digitally authenticatable today, whereas others may need to be physically presented in person to avoid spoofing.\footnote{Some national IDs hold near-field communication (NFC) readable, cryptographically signed information like name, address, and date of birth as well as face images. For a description of NFC in ePassports, see \cite{noauthor_epassport_nodate}. This allows a holder to demonstrably confirm their ownership of the physical card in the digital domain, unlike ID cards in countries like the United States (including some REAL ID-compliant documents), which often rely upon sending a (spoofable) photo or video for digital authentication (depending on the permissions of the verifying party).} One challenge may be how to authenticate IDs from different jurisdictions, in a relatively uniform manner. Usefully, there are already global standards for some forms of ID (such as the United Nations’ oversight of ePassport standards), with public key infrastructure built for authenticating these; third-parties can then explore layering zero-knowledge technology atop the documents.\footnote{For more on ePassports, see \cite{noauthor_epassport_nodate}. For an overview of how various systems come together to enable identification for cross-border movement, see \cite{dermarkar_facilitation_2022}. For an example of layering zero-knowledge technologies atop these documents, see \cite{noauthor_zk-passportproof--passport_2024}.}

Biometrics are another method for limiting the number of credentials per person from an issuer; these depend on measuring a part of a person (e.g., palm, iris, fingerprint) that is persistent and unique to them, and then checking against matches from other applicants.\footnote{Different biometric approaches vary in their accuracy---groups such as NIST have released de-identified datasets of biometric indicators to aid with improving the accuracy of biometric methods \cite{nist_releases_2019}.} One challenge for biometric systems is affirming the integrity of hardware devices used to collect measurements.\footnote{It can be difficult to verify that proprietary hardware designed for collecting biometric information is free from backdoors \cite{buterin_what_2023}. Generalized hardware, on the other hand, might be insufficiently sensitive to prevent forms of presentation attacks \cite{gartenberg_hacker_2017}. One approach is for a builder of proprietary hardware---such as Worldcoin, which has built Orb devices to facilitate iris recognition---to release its schematics \cite{worldcoin_hardware_2023}. This can increase transparency and might ultimately help other groups to manufacture such devices, but does not fully resolve the challenge of hardware integrity.} Because biometric-based systems inherently involve the processing or storing of sensitive information about people, it is important the data be handled in a transparent and privacy-preserving manner (e.g., practices like hashing and encrypting the data).\footnote{Some biometric systems try to limit the risks of storing biometric information by not storing other sensitive data, but there are always difficulties in determining what sensitive information can be inferred through correlations in non-sensitive information. For instance, India’s Aadhaar, a large-scale biometric identity system, states that they do not store sensitive information such as religion or caste alongside biometrics \cite{noauthor_how_nodate}, but there are criticisms of this approach based on the possibility of linking databases \cite{sadhya_critical_2024}.}  Many privacy and civil liberties groups object to biometric systems, particularly without strong checks against abuse.\footnote{See, for instance, criticism of direct uses of biometrics by governments and via private organizations in partnership with the government \cite{aclu_fight_2023, aclu_three_2022}.  At the same time, regulators may be concerned about the legality of how private actors operate biometric PHC systems---Worldcoin, for instance, has been investigated in a number of jurisdictions \cite{worldcoinhongkong,worldcoinspain, njenga_global_2024}, and some of these bans have since been overturned \cite{worldcoinkenya}.} Some researchers have proposed methods for decreasing the risks of abuse, such as by decentralizing the storage of biometric information \cite{acquah_securing_2020,othman_method_2018, worldcoin_hardware_2023}. One commonly cited potential benefit of biometric-based systems is in their near universality---nearly all people will have biometric indicators (like fingerprints) to enroll in a PHC system,\footnote{Even in cases of near-universal availability of a biometric method, an issuer will need to consider potential biases in accuracy rates across demographic groups \cite{buolamwini_gender_2018,fang_demographic_2021}.} whereas fewer may have a certain government ID\footnote{As noted in \Cref{sec:risks}, there are around 850 million people worldwide who do not currently hold official identity documents \cite{world_people_2023}.} or be well-connected in a social graph.

Web-of-Trust (WoT) is yet another method for limiting the number of credentials per person: issuers depend on social graph analysis to establish that a person is real (e.g., ``Are they vouched for by a person who is strongly believed to be real?'') and has not already received a credential. Often, these systems are ``seeded'' with initial trusted users from some other method of trust (e.g., existing IDs, biometrics, or in-person ``pseudonym parties'') \cite{ford_offline_2008}. Someone without a seed-authenticated profile can become authenticated through sufficient vouches from already-trusted parties. While WoT can be a strong method for validating that a person exists, these systems can struggle to confirm uniqueness: A person may be able to get multiple credentials from multiple distinct social circles.\footnote{Determining whether a person has already received a credential can be a more difficult challenge for WoT-based methods. One way to assess this may be to analyze ``Does their social graph look similar to cases where people have been found to have improperly obtained multiple credentials?'', though this will invariably be prone to errors. For a review of WoT-based approaches, see \cite{siddarth_who_2020}.} The most fraud-resistant mechanisms for WoT may involve layering another method, such as biometrics, atop WoT, though in a way such that users still feel comfortable with privacy assurances.\footnote{For instance, one option for layering simple biometrics atop WoT may include people having a private profile picture, which is viewable only to a small subset of trusted contacts but can still be compared against other private profile pictures in a privacy-preserving way. Use of any biometrics may undermine a system’s reasons for having initially selected WoT as a method, however.} To aid in enforcing one-per-person credential limits, WoT issuers may also consider other signals beyond one’s social graph—such as geographic consistency, transaction patterns, and device usage (all commonly used in modern anti-fraud frameworks today).

\subsubsection*{\underline{Expiry or regular re-authentication}}
\label{ssapx:expiry}

Importantly, issuers must protect the one-credential-per-person requirement through secure re-authentication (checking whether the credential is held by the party to whom it was issued) and/or time-bounded expiry (limits on a credential’s lifespan, which impact one actor’s ability to maintain multiple over time). As re-authentication can be difficult to achieve in a privacy-preserving manner,\footnote{For an overview of authentication methods and their trade-offs, see \cite{shostack_threat_2014}.} it may be easier for some issuers to simply set tight expiration limits on credentials.

\subsection{\textit{Methods to achieve unlinkable pseudonymity in practice}}
\label{sapx:methods_unlinkable}

\subsubsection*{\underline{Minimal identifying information stored during enrollment}}
\label{ssapx:minimal_information}

The amount of information that an issuer learns and stores about an enrollee will depend on the method by which the issuer achieves its per-person credential limit, and how the issuer chooses to handle lost, stolen, or otherwise compromised\footnote{While it is difficult to prevent the transfer of PHCs, there are methods for achieving some forms of non-transferability for anonymous credentials \cite{camenisch_efficient_2001, lysyanskaya_pseudonym_2000} and other digital assets more generally \cite{dwork_digital_1996}.} credentials. Specifically, consider a scenario in which the issuer stores no information about the user after enrollment. This preserves privacy, but if the user loses or loses control of their PHC, how does the issuer determine which PHC needs to be recovered or revoked? Different systems resolve this trade-off between privacy and ease of recovery and revocability in different ways.

It is possible that once a PHC is issued, the issuer stores no identifying information at all about a credential holder.\footnote{Note that PHCs are thus far more private than ``real name'' policies \cite{york_facebooks_2014}.} For instance, even if using an existing government ID as the basis for issuing a credential, the issuer can record that an ID number has been used for this purpose without recording which PHC was issued to said ID-holder. This process still allows the issuer to check for duplicated sign ups, enabling the credential limit.\footnote{One must also be careful to avoid other forms of correlation---for instance, if an issuer can link a specific person to their PHC through a field like time-of-issuance.}

An issuer that does not store any identifying information must carefully design their recovery and revocation process. There is a substantial body of literature in cryptography detailing the challenge of recovery for anonymous credentials, which have much in common with PHCs \cite{camenisch_dynamic_2002}. Beyond the partial solutions suggested in that literature, there are other methods that would violate some of the strict requirements of anonymous credentials, but may be suitable for use in PHC systems. For instance, issuers could use methods like back-up codes, security questions (which must be chosen to not be identifying), and hardware tokens.\footnote{For an overview of authentication methods and their tradeoffs, see \cite{shostack_threat_2014}.} When the issuer does store identifying information for use in recovery and revocation, it should be encrypted.\footnote{For a discussion of how to protect sensitive data, see \cite{mccallister_guide_2010}.}

Furthermore, issuers may decide that a less-than-foolproof recovery and revocation process is tolerable---worth the cost for the gains in privacy. In such cases, the issuer could make other design decisions that mitigate the harms from a sometimes-faulty recovery and revocation process. For instance, as discussed above, setting a shorter length of time during which a credential is valid can mitigate the harms when a holder has lost their credential. They need only wait a short period before they would need to renew their credential anyway. Issuers can also encourage credential holders to follow best practices to safeguard their credential and avoid loss or theft in the first place. Furthermore, the presence of other PHC issuers in the broader ecosystem may mitigate the harms from losing a credential—the holder with a lost credential can acquire a new one from another issuer.

\subsubsection*{\underline{Minimal disclosure during usage and unlinkable pseudonymity}}
\label{ssapx:minimal_disclosure}

By drawing on public key cryptography,\footnote{For foundational developments in public-key cryptography, see \cite{diffie_new_1976, rivest_method_1978, miller_use_1986}. For a textbook treatment, see \cite{menezes_handbook_1997}.} zero-knowledge proofs, and mechanisms like cryptographic nullifiers, PHC systems can achieve minimal disclosure and unlinkable pseudonymity. There are a wide range of protocols studied in theory and applied in practice that use these cryptographic building blocks to satisfy requirements analogous to the ones we have defined.\footnote{There are several decades worth of research into the affordances of various systems for issuing anonymous credentials. For two prominent implementations of anonymous credentials, Idemix \cite{camenisch_design_2002} and U-Prove \cite{paquin_u-prove_2013} offer practical implementations for privacy-preserving authentication.} The following is intended only as a brief sketch of how these tools could be employed in PHC systems.

Public key cryptography is likely to be a foundational building block for a PHC system, because it allows an issuer to keep track of valid credentials in a privacy-preserving way. For instance, a PHC issuer could maintain a list of public keys (each related to a valid credential), each of which has a paired secret private key. When a new person successfully enrolls in the PHC system, the issuer lets this person add exactly one public key to the list of valid keys---the private key is known only to the user enrolling.

How does a user prove they are the holder of a valid key from the issuer, without revealing which key they hold?\footnote{Note that unlinkable pseudonymity would not be achieved if the service provider were to learn which key the user holds---if so, the key could be used to track the user across different service providers.} One way to do this is with zero-knowledge proofs.\footnote{Researchers in both academia and industry are actively working on the design of zero-knowledge proof protocols—many designs are efficient enough to be deployed in the real world. Different classes of zero-knowledge protocols are optimized for different requirements.  Recently, there have been significant efforts focused on developing \emph{generic zero-knowledge}, which can be used to prove almost any statement of interest. When bandwidth concerns dominate, zk-SNARKs \cite{bitansky2012extractable} and zk-STARKs \cite{ben2018scalable} tend to be the best generic zero-knowledge choices because of their short proof sizes. MPC-in-the-head and Vector-OLE approaches may have better performance profiles when system designers instead aim to minimize the time required to generate a proof. While flexible, generic zero-knowledge proofs tend to be less concretely efficient than bespoke zero-knowledge proofs optimized to prove specific statements. A notable example of a bespoke zero-knowledge proof that could be of use in personhood credential systems is a {ring signature}  \cite{rivest_how_2001}, which allow provers to sign messages while only disclosing that they are a member of some publicly known group, such as the group of valid credential holders, but not which specific member of the group.} As discussed at several points in the paper, zero-knowledge proofs are cryptographic protocols that enable a “prover” to convince a “verifier” of a statement’s truth, without revealing any additional information beyond the validity of the statement. In the case of PHC usage, the user who holds a PHC is the prover and a service provider is the verifier. The user proves to the service provider that the statement “I hold a valid PHC” is true, without revealing which PHC---for instance, by proving “I hold a secret private key that pairs with some public key on the issuer’s list.” 

Beyond verifying that the user possesses a valid PHC, service providers may need a method that allows them to check whether a PHC has been used with their service before---they need to create a service-specific ``pseudonym'' for the credential holder. (Indeed, this ability is what enables service providers to rate-limit the use of their service with a PHC, a foundational motivation for building PHC systems.) One approach to creating service-specific pseudonyms employs cryptographic nullifiers---paralleling solutions in e-cash systems that were designed to prevent double-spending, where unique identifiers ensure a digital coin is not spent more than once  \cite{chaum_blind_1983, goldwasser_untraceable_1990}.\footnote{For example implementations of cryptographic nullifiers, see \cite{gabizon_how_2016, sasson_zerocash_2014}.} In the context of PHCs, when a user interacts with a service, they compute a unique nullifier---a number or string---based on their credential and the service’s identity. The service can store these nullifiers to track whether a credential has been used before, without learning any information about the user’s credential, as the service cannot ``undo'' the computation to derive the specific credential. Crucially, while the service can determine if a credential has been used in a specific context, these nullifiers are context-specific and do not allow linking the user’s activities across different services or contexts.

\subsection{\textit{Issuers’ incentives and governance}}
\label{sapx:incentives_governance}

As noted throughout the paper, there are many possible issuers of personhood credentials, including trusted institutions like governments, nonprofits, consortia, and private companies. It is important to consider the incentives of a potential PHC issuer: A democratic nation issuing PHCs has different inherent incentives than an authoritarian one, and both have different inherent incentives than a nonprofit.

Ideally, a PHC issuer’s incentives are aligned with the majority of its users’, i.e., the issuer would like to ensure that the system runs with high integrity and smoothly facilitates trustworthy digital activity as intended. One approach to reinforcing these incentives might be to set up a global consortium of participating organizations and governments, following a multistakeholder governance model. An analog today might be the Internet Corporation for Assigned Names and Numbers (ICANN), which coordinates the maintenance and procedures of several databases related to the namespaces and numerical identifiers of the Internet.\footnote{ICANN’s core governance components are a multistakeholder policy development process rolling up to a policy-making council, additional supporting organizations, and a governmental advisory committee, as well as periodic public meetings intended to field broader popular input. For a detailed study of ICANN’s accountability framework and legal cases related to ICANN’s review process, see \cite{petillion_competing_2017}.}

A related question is how an issuer decides which use-cases their personhood credential supports. For instance, an issuer might wish to have a registration process for service providers by which they can be approved and gain access to a verification protocol,\footnote{In recent years, there have been many examples of technology platforms deciding to exclude certain types of use-cases from their protocol---for instance, in the case of the credit card networks and online pornography \cite{goodwin_mastercard_2020, fung_why_2021}. Likewise, some leading identification systems are not universal in their support of use-cases today: For instance, India’s Aadhaar has an application process for certain entities to gain access to authenticating Aadhaar numbers, given a number of security requirements entailed for authenticators \cite{UIDAI_requesting_nodate}.} though this may be in tension with the requirement that issuers not gain access to usage data for any particular person’s personhood credential.\footnote{It is worth noting that there are two distinct questions here: 1) Which service providers can gain access to a PHC system’s verification protocol, and for what services?, and 2) When a particular verification happens, what information is made available to the issuer? By our requirements, a PHC issuer should not learn about any particular usage via verification---but the issuer might be able to limit who can access the verification protocol, without violating that requirement.} A more interoperable PHC may have benefits of reduced friction for users, who would not need to go through issuance for many different applications---though this may also increase the risk to users if they lose control of their personhood credential, and may increase the risks to privacy and civil liberties associated with a centralized issuer, detailed in \Cref{apx:ecosystem}. Notably, if there are multiple issuers in a PHC ecosystem, then users can select into credential issuers with their desired level of interoperability.

\appsection{Ecosystem design and management}
\label{apx:ecosystem}

Recall that a personhood credentialing ecosystem has two goals: to reduce the harms of scaled deception while protecting privacy and civil liberties. We begin with a discussion of how an ecosystem in which each person can hold a bounded number of credentials (greater than one but not too many) resolves tensions in key design goals. Then we discuss service provider and issuer choices that must be considered in an ecosystem in which users hold may more than one credential.  

\subsection{Tensions in design goals can be resolved with bounded credentials}

Here, we highlight some inherent tensions between these ecosystem design goals through two extremes, which we disfavor: one in which people can acquire an unlimited number of credentials, and one in which people can acquire only one credential from a single issuer.

Through these extremes, we illustrate that an ecosystem with bounded credentials---dependent upon having a per-person credential limit from each issuer---may best resolve these tensions. Then, we discuss how service providers might protect the integrity of their services when each person may have more than one (but not too many) PHC.

\subsubsection*{\textit{Unlimited credentials}}
\label{sapx:unlimited_credentials}

First, consider the extreme in which issuers have no credential limits; a person can obtain many credentials from an issuer without difficulty.\footnote{In practice, ``not having credential limits'' may be a spectrum more than a binary. For instance, we expect there are significant differences between issuers that officially allow holders to have unlimited unlinked credentials, and issuers that would prefer holders to have few but do not have very effective enforcement of this.} In this case, when there are unlimited credentials per person, it is hard to use these credentials to address the harms of scalable deceptive behavior. For instance, service providers cannot enforce per-person rate limits to stop repeated abuse.

On the other hand, such an ecosystem may have fewer threats to user privacy and civil liberties: Easy acquisition of credentials reduces the power of any single issuer to control the flow of credentials or their uses. Further, such an ecosystem may require less sensitive data from users: The issuer does not need to process or store any identifying information in order to check for duplication before dispensing a credential; they only need to check whether the user in question is a person.\footnote{Such systems might also have lower stakes in the case of a holder losing their credential: If there is no limit on the number of credentials someone can obtain, a holder can be issued with a new one with fewer complications.}

\subsubsection*{\textit{One credential}}
\label{sapx:one_global_credential}

Next, consider the opposite extreme, in which there is only one issuer of personhood credentials in the ecosystem, and each person can only obtain one personhood credential from the issuer. This approach makes it more difficult for bad actors to obtain multiple credentials for use in scaled deception, though with significant challenges to privacy and civil liberties.

One challenge is that participants will have less choice: Even if they object to the issuer’s method of ensuring that people do not already have a credential, their only choice is to concede or to be excluded from participation in the ecosystem. We are concerned about these dynamics and how they may concentrate power, particularly with issuers that many people have no pre-existing reasons to trust.\footnote{There are many reasons that people might be hesitant about trusting an issuer: for instance, if they have had little contact with the issuer previously, or if the issuer is not clearly accountable in some form to the holders.}\footnote{One further challenge of a single issuer is that revocation and recovery processes will have higher stakes: With only a single issuer, people would not have the ability to merely obtain a credential from another issuer instead. This might also increase an issuer’s need to collect identifying information about holders so that the issuer knows which PHC to invalidate when they issue a new PHC to this person. As long as the issuer cannot also track PHC \textit{usage} (such as via collusion with service providers), the practical benefits of preserving a link between PHCs and some aspects of a person’s identity at enrollment may still outweigh the privacy costs.} More generally, having multiple issuers in an ecosystem could help encourage improvements in quality and demonstrated trustworthiness.

\subsubsection*{\textit{Bounded credentials}}
\label{sapx:bounded_credentials}

Given the inherent tensions highlighted by the extremes of ecosystems with unlimited credentials and ecosystems with one credential, we advocate for ecosystems that aim to bound the number of credentials obtainable by each person. This approach balances between the twin goals of addressing scaled deception and protecting privacy and civil liberties.

When each person can only obtain a bounded number of credentials in the overall ecosystem, it is feasible to enforce per-person rate limits and account creation limits.\footnote{Though it is feasible to enforce per-person rate limits, an ecosystem with a bounded number of credentials might not be able to strictly achieve a \textit{one}-per-person limit: For instance, in an ecosystem with three well-trusted PHC issuers, a service provider might decide to allow PHCs from each of these---and thus, even if the service provider allows only one verified account per PHC, someone could obtain up to three accounts by using one PHC from each issuer.}

The bounded network also reduces the number of credentials in circulation, thus decreasing the supply available for sale, transfer, and theft.\footnote{Attackers face greater challenges in obtaining additional credentials if issuers do enforce per-person limits: (1) Attackers would need to manipulate the primary issuance process, which is more complex and costly because the issuer checks if they already hold a credential. (2) Purchasing credentials from a secondary market would become more expensive and difficult because there is a limited supply, and legitimate users have a higher opportunity cost for selling their sole credential. Thus, issuers adopting per-person credential limits help to enhance security against attackers without significantly hindering legitimate users from obtaining their individual credentials.}

\subsection{\textit{Choices for service providers and issuers in an ecosystem with multiple issuers}}
\label{sapx:choices_for_providers}

When an ecosystem has multiple PHC issuers---even if each has a credential limit and thus the ecosystem has a bounded number of credentials on the whole---there are complex trade-offs that both service providers and issuers will need to consider.

For issuers, they may face decisions of whether---and in what forms---to try to cooperate with other issuers to deduplicate credential holders, who may otherwise have more than one PHC. Given that some users may have chosen a particular issuer to avoid having any relationship with a different particular issuer, under what circumstances is this appropriate?

For a service provider, decisions will often center around which issuers’ PHCs to accept and how these PHC options relate to one another. Trade-offs will often entail balancing the service’s ability to reduce scaled deception by ensuring each person has only one credential, compared with the implications of having fewer choices for its users (such as users feeling pressure to use a credential that is difficult for them to obtain, or the increased concentration of power in a chosen issuer, which creates challenges as discussed in \Cref{sec:risks}).

There are no perfect solutions to these challenges; having multiple issuers of PHCs necessarily will involve choices of which to accept, which may affect services’ abilities to rate-limit activities that are potentially deceptive. We encourage significantly more research on methods to balance these desiderata in an ecosystem. Below, we briefly discuss a number of considerations.

One principle that service providers might consider is to aim for inclusivity—trying to increase the number of their users who have access to an accepted PHC. For example, if most of a service’s users live in a state that has an associated PHC backed by its state ID, the service can look to complement this by accepting a PHC that is more common among the portion who do not live in this state.

A second principle that service providers might consider is to aim for integrity---trying to reduce the number of deceptive users who would have easy access to a large number of PHCs.\footnote{If a user is limited to only a few PHCs---say, three instead of one---this may not be ideal for reducing deceptive behavior, but it is significantly less challenging than if a user can obtain 100.} This may be achievable, for instance, if there are PHC systems that have non-overlapping eligibility criteria: Two PHC issuers may each require that holders of their PHC reside within their state. Because it would be difficult for a deceptive user to successfully obtain each of these PHCs at once, the service provider should expect that fewer will succeed at evading the service’s per-person intentions.

We also consider these questions from a PHC issuer’s perspective; are there actions that an issuer can take that would reduce the risk of duplicative use with a service, while still allowing people to have choice among PHCs?

One approach may be possible if issuers rely on the same underlying method to determine who has already received a PHC from them. For instance, if issuers each rely upon a particular state ID card, it might be possible for these issuers to coordinate on a sufficiently private mechanism for checking whether the card has previously been used with the other, prior to issuing a PHC directly to the holder. 

We expect that these challenges will not be trivial to solve and may involve difficult trade-offs. Nonetheless, we believe that the benefits of an ecosystem with multiple issuers can be quite large---particularly for privacy and civil liberties---and so encourage further research and exploration.

\end{document}